\documentclass[prd, english,reprint, longbibliography, superscriptaddress, breaklinks=true, showkeys, showpacs=false, nofootinbib]{revtex4-2}
\usepackage[T1]{fontenc}
\usepackage[utf8]{inputenc}
\usepackage{color}
\usepackage{babel}
\usepackage{amsmath}
\usepackage{amssymb}
\usepackage{subfigure}
\usepackage{graphicx}
\usepackage{physics} 
\usepackage{dsfont}
\usepackage{hyperref}
\usepackage{float}

\newcommand{\vx}{\textbf{x}}

\begin{document}

\title{Heat distribution of quantum fields interacting with Unruh-DeWitt detectors}

\author{Marcos L. W. Basso\href{https://orcid.org/0000-0001-5456-7772}{\includegraphics[scale=0.05]{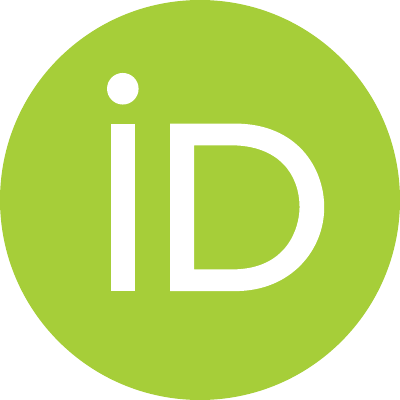}}}
\email{marcoslwbasso@hotmail.com}
\affiliation{Departamento de Matemática Aplicada, Universidade Estadual de Campinas, 13083-859 Campinas, S\~ao Paulo, Brazil}

\author{Alberto Saa\href{https://orcid.org/0000-0003-1520-4076}{\includegraphics[scale=0.05]{orcidid.pdf}}}
\email{asaa@ime.unicamp.br}
\affiliation{Departamento de Matemática Aplicada, Universidade Estadual de Campinas, 13083-859 Campinas, S\~ao Paulo, Brazil}
\begin{abstract}
We introduce an operational framework to characterize the statistics of heat exchange between a quantum field and an Unruh–DeWitt particle detector in Minkowski spacetime. Building on an interferometric protocol, we define the characteristic function of heat without relying on projective measurements, ensuring consistency with relativistic causality. Working perturbatively, we obtain general expressions for the characteristic function and the corresponding heat distribution. In the vacuum, the statistics coincides with the low-intensity expansion of a unidirectional Poisson law, while for Kubo-Martin-Schwinger (KMS) states it acquires a bidirectional Poisson structure up to second order in perturbation theory. We study how fluctuation relations emerge from the interplay between detector properties and field correlations, and identify regimes where these relations reduce to well-known fluctuation relations for heat exchange. In particular, we discuss an analytical derivation for the entropic Landauer inequality as a direct specialization of the algebraic entropy-balance theorem of Jak\v{s}i\'c and Pillet and verify it numerically within the perturbative detector–field interaction. We further compare its tightness with that of a complementary Landauer bound derived from the full counting statistics of heat, analyzing how their relative performance depends on the detector's initial state and interaction time. We also formulate the first and second laws for the explicitly switched detector–field dynamics, identifying the external switching work that reconciles the field-energy gain with the detector-energy change. Finally, in the gapless and delta-switching regimes, where the dynamics can be treated non-perturbatively, the characteristic function coincides with that of a bidirectional Poisson process, leading to an exact fluctuation theorem determined by the KMS condition. These results provide a consistent framework for heat statistics in relativistic quantum field theory and clarify the emergence of fluctuation relations in this context.
\end{abstract}

%\keywords{Heat distribution; Fluctuation relations; Static curved spacetimes}

\maketitle

%%%%%%%%%%%%%%%%%%%%%%%%%%%%%%%%%%%%%%%%%%%%%%%%%%%%%%%%%%%%
\section{Introduction} 

Over the past few decades, quantum information has emerged as a major development in physics, extending classical information concepts into the quantum regime, driven by both foundational advances in quantum theory and improved experimental control at microscopic scales~\cite{Simmons, Adesso}. In this context, it has become clear in recent years that one of its central challenges is to understand its relationship with two other fundamental pillars of physics: quantum field theory (QFT) and general relativity (GR)~\cite{Preskill, Peres2004}. In particular, relativistic quantum information studies how information-theoretic protocols arise from local interactions with quantum fields and how they are affected by relativistic effects~\cite{Alsing2003, Hu2012, Landulfo2016, Tjoa2022a, Perche2023, Kasprzak2025}. This framework includes both relativistic extensions of standard protocols and intrinsically field-theoretic phenomena, such as entanglement harvesting from the vacuum~\cite{Valentini, Resnik, Pozas2015, Perche2024}, as well as protocols like quantum collect calling~\cite{Jonsson2015} and quantum energy teleportation~\cite{Hotta2008}.

Information theory has also reshaped statistical mechanics and thermodynamics. An early and influential contribution in this direction is Jaynes’ principle of maximum entropy~\cite{Jaynes1957a, Jaynes1957b}, which reformulates statistical mechanics as an inference problem based on incomplete information. Within this perspective, thermodynamic ensembles arise from maximizing entropy subject to known constraints, thereby establishing a direct link between information and physical entropy. Another key insight is given by Landauer's principle~\cite{Landauer}, which states that the erasure of one bit of information necessarily entails an increase in thermodynamic entropy. This idea played a central role in Bennett’s resolution of Maxwell's demon paradox~\cite{Bennett, Plenio, Chaves2025}, clarifying how information processing reconciles apparent violations of the second law. Since then, several approaches and physical frameworks have been developed to derive and generalize the Landauer bound, ranging from information-theoretic arguments to dynamical and operational formulations in both classical and quantum settings~\cite{Esposito2010, Hilt, Reeb2014, Goold2015, Campbell2017, Guarnieri2017, Taranto2023, JaksicPillet2014}.

With the advent of quantum information theory, these connections have gained renewed significance~\cite{Goold2016a}. In particular, they have led to the development of quantum thermodynamics, a field that aims to generalize thermodynamic principles to quantum systems while accounting for uniquely quantum features such as coherence, entanglement, and fluctuations at small scales~\cite{Vinjanampathy, Alicki}. At the same time, nonequilibrium thermodynamics has undergone substantial progress, especially in the study of systems driven far from equilibrium, where linear response theory no longer applies. A major outcome of these developments is the formulation of fluctuation theorems, such as those by Jarzynski and Crooks, which establish exact relations between equilibrium and nonequilibrium thermodynamic quantities~\cite{Evans, Jarzynski1997, Crooks1998, Crooks1999, Jarzynski2004, Esposito2009, Campisi2011, Jarzynski11, Seifert2012, Landi2021}. These results provide a quantitative characterization of irreversibility, showing that processes with negative entropy production—corresponding to apparent violations of the arrow of time—are exponentially suppressed. 

Building on these developments, extending fluctuation theorems to quantum fields in curved spacetimes remains a largely open problem. Such an extension faces both conceptual and technical challenges. In generic curved geometries, the absence of global symmetries and of a preferred notion of time prevents a straightforward generalization of standard notions, such as equilibrium states. Progress in this direction has been achieved in flat-spacetime quantum field theory. In particular, Jarzynski equality was formulated for driven quantum field theories using a two-point measurement (TPM) scheme applied to time-dependent field dynamics~\cite{Bartolotta}. However, the TPM scheme, which underlies many formulations in quantum thermodynamics, becomes problematic when directly applied to relativistic quantum fields. In particular, the use of projective measurements in this context generically leads to violations of relativistic locality, as it can enable superluminal signalling between spacelike separated regions~\cite{Sorkin1993, Borsten, Bostelmann, Anastopoulos2022, Papageorgiou}. These difficulties indicate that alternative operational frameworks are required. Recent developments in relativistic quantum information and quantum field theory support this perspective, suggesting that localized measurement models~\cite{Fewster2020}—such as those based on Unruh–DeWitt (UDW) detectors~\cite{Unruh1976, DeWitt1979, Schlicht04, Louko06, Louko08, Satz2007, Ramon21, Tjoa22, Ramon23, Polo}—offer a consistent way to define observables while respecting locality.

In parallel, significant progress has been made in understanding the thermal properties of quantum fields in both flat and curved spacetimes~\cite{Haag1996, Sewell1982, Takagi1986, Kay1991}, as well as the thermalization behavior of UDW detectors~\cite{Fewster, Garay, Aubry, Good2020, Perche2021}. Related developments have also explored the formulation of relativistic quantum thermodynamics for moving detectors~\cite{Papadatos}. Regarding nonequilibrium phenomena in curved spacetime, Iso \textit{et al.}~\cite{Iso2011} analyzed fluctuations at black hole horizons by combining the Jarzynski equality with the generalized second law~\cite{Bekenstein74, Wall2012}. More recently, a detailed fluctuation theorem was formulated for localized nonrelativistic quantum systems in curved spacetime~\cite{Basso2025}. Analogous relations have also been obtained for classical stochastic systems~\cite{Cai25}.

In Ref.~\cite{Ortega}, the first consistent definition of a work distribution for quantum fields coupled to a UDW detector in Minkowski spacetime was introduced, inspired by Ramsey interferometry~\cite{Dorner2013, Mazzola2013}. In this context, the Ramsey interferometric scheme requires a specific form of a controlled unitary between the field and the detector in order to encode the characteristic function of work in the detector's density matrix. Some of its main properties were analyzed, including work fluctuation relations and the connection between fluctuations and expectation values. Building upon this framework, Ref.~\cite{Costa2026} advances the program by generalizing the protocol to the case of a quantum scalar field in a globally hyperbolic and static curved spacetime. From a different perspective, Ref.~\cite{Hong2025} employs a Schwinger-Keldysh contour in an effective field theory to study work statistics in nonequilibrium many-body quantum systems. Moreover, in Ref.~\cite{Bonfill}, the authors show that, in Minkowski spacetime, for the class of spacetime-localized unitary processes on a quantum field implementing the protocol of Ref.~\cite{Ortega}, the average internal energy change coincides with the average work and that the variances of internal energy and work also agree. However, in more general situations, the first law also involves a contribution associated with heat, which is not captured by work statistics alone. This indicates that a complete thermodynamic description requires an operational notion of heat and its corresponding statistical properties. 

Motivated by these considerations, in this work we extend the interferometric approach to define a heat distribution for a quantum field interacting with a UDW detector, based on Refs.~\cite{Goold2014, Peterson}, which introduced heat distributions using Ramsey interferometry, although the corresponding protocol differs from those employed in the characterization of work. This construction provides a natural counterpart to the work distribution framework and opens the way to investigate heat statistics and fluctuation relations associated with heat exchange, as well as to explore fundamental aspects such as the Landauer bound in relativistic quantum field settings, which we pursue in the present work. Hence, a second objective of this work is to present a Landauer bound formulation  compatible with an infinitely extended quantum field. We apply the algebraic entropy-balance theorem of
Jak\v{s}i\'c and Pillet~\cite{JaksicPillet2014} to the finite-dimensional detector coupled to a Kubo-Martin-Schwinger (KMS) reservoir, using Araki relative entropy~\cite{Araki1976,Araki1977}. This gives the analytically entropic inequality, subject to the regularity condition on the unitary interaction stated in Sect.~\ref{sec:landauer-bounds}.  We perform a numerical verification to compare this bound with the full-counting-statistics bound, displaying how their relative tightness depends on the detector state and interaction time. It is worth mentioning that a related study of Landauer's principle in a quantum field theoretic context was presented in~\cite{Xu2022}, where a qubit interacts perturbatively with a scalar field confined in a cavity. This setup provides a simple and well-defined framework in which the exchange of energy between the qubit and the field, as well as the variation of the qubit's entropy, can be numerically evaluated. In contrast, in our work we do not assume a cavity and instead consider a field interacting with a qubit in a localized region of Minkowski spacetime, allowing for a more direct connection with standard QFT scenarios.

The remainder of this paper is organized as follows. In Sec.~\ref{Sec:II}, we begin by reviewing the quantization of a scalar field and introducing Unruh--DeWitt particle detectors as localized probes. We then present the interferometric protocol used to define the characteristic function of heat for a quantum field interacting with a UDW detector. Within the same section, we discuss two formulations of the Landauer bound: the entropic formulation, established through the algebraic entropy-balance theorem, and the bound derived from the full counting statistics and the associated fluctuation relation. We further formulate the first and second laws for the explicitly switched detector–field dynamics, relating the switching work to the energy balance and to a nonequilibrium free-energy bound. In Sec.~\ref{Sec:III}, we present the main results within second-order perturbation theory, deriving the heat distribution and analyzing its properties for both vacuum and KMS field states. We also investigate the emergence of fluctuation relations and their dependence on the detector's initial state and interaction time. Finally, we evaluate the two Landauer bounds within the perturbative UDW model and compare their numerical tightness as a function of the detector preparation and interaction time. In Sec.~\ref{Sec:IV}, we investigate the gapless and delta-switching regimes, where a non-perturbative treatment is possible and exact fluctuation relations can be established. Finally, in Sec.~\ref{Sec:V}, we summarize our results and discuss possible extensions.  We employ the metric signature ($-,+,+,+$) and Planck units throughout the manuscript.

\section{Heat distribution for quantum fields}
\label{Sec:II}
In this section, we briefly review the quantization of a real scalar field in Minkowski spacetime within the algebraic framework~\cite{Haag1996}. We then introduce UDW particle detectors as local probes of quantum fields, which provide an operational way to access expectation values through localized interactions. This setting will serve as the basis for the interferometric protocol used to define the heat distribution, and is included to keep the presentation self-contained. Before introducing the interferometric protocol for the quantum field, we review the definition of heat distributions based on the TPM scheme for systems with finite degrees of freedom, and discuss why this approach is not well suited for quantum fields. Finally, we discuss the Landauer bound as formulated in~\cite{Reeb2014, JaksicPillet2014}, as well as the non-equilibrium Landauer bound derived from fluctuation relations~\cite{Goold2015}. In the following sections, these bounds will be investigated in a relativistic setting using the heat distribution defined through the interferometric protocol.

\subsection{Setting the stage} 
\label{sec:setting}
Let us begin by considering Minkowski spacetime $(\mathcal{M}, \eta)$, with
$\eta = - dt \otimes dt + \delta_{ij} dx^i \otimes dx^j,$ where $(t,\textbf{x})$ are the inertial coordinates that cover Minkowski spacetime and label events $\mathsf{x} \in \mathcal{M}$. The quantization procedure for a real scalar field can be formulated within the algebraic approach by associating to each test function $f \in C_0^\infty(\mathcal{M})$ an element $\hat{\phi}(f)$ of a unital $*$-algebra $\mathcal{A}_{\phi}(\mathcal{M})$. This defines a linear map
\begin{equation}
\hat{\phi}: C_0^\infty(\mathcal{M}) \to \mathcal{A}_{\phi}(\mathcal{M}), \qquad f \mapsto \hat{\phi}(f),
\end{equation}
which encodes the quantum field as an operator-valued distribution. Formally, one may write
\begin{equation}
\hat{\phi}(f) = \int d^4 \mathsf{x}  f(\mathsf{x}) \hat{\phi}(\mathsf{x}).
\end{equation}

The field algebra is defined by imposing the following conditions. First, hermiticity implies that $\hat{\phi}(f)^\dagger = \hat{\phi}(f^*),$ for all $f \in C_0^\infty(\mathcal{M})$. Second, the Klein–Gordon equation holds in the distributional sense, i.e.,
\begin{equation}
\hat{\phi}\left((\nabla_a\nabla^a - m^2)f\right) = 0,
\end{equation}
where $\nabla_a$ is the covariant derivative compatible with Minkowski metric $\eta$. Third, the canonical commutation relations (CCR) are given by
\begin{equation}
[\hat{\phi}(f_1), \hat{\phi}(f_2)] = i E(f_1,f_2)\hat{\mathbb{I}},
\end{equation}
for all $f_1,f_2 \in C_0^\infty(\mathbb{R}^4)$, where $E$ denotes the causal propagator
\begin{equation}
E(f_1,f_2) = \int_{\mathcal{M}} d^4 \mathsf{x} f_1(\mathsf{x}) E f_2(\mathsf{x}), \label{eq:Causalprop}
\end{equation}
with $E = G^{\text{ret}} - G^{\text{adv}}$ being the difference between the retarded and advanced Green operators of the Klein–Gordon operator. Finally, let $\Sigma \subset \mathcal{M}$ be a Cauchy surface and $\mathcal{O}$ a fixed open neighbourhood of $\Sigma$. According to the time-slice axiom, the field algebra $\mathcal{A}_{\phi}(\mathcal{M})$ is generated by the unit element $\hat{\mathbb{I}}$ and the smeared field operators $\hat{\phi}(f)$ with $f \in C_0^\infty(\mathcal{M})$ and $\mathrm{supp}(f) \subset \mathcal{O}$. In this sense, the algebra of observables in the entire spacetime is fully determined by the field operators localized in any neighbourhood of a Cauchy surface.

In Minkowski spacetime, the presence of a global timelike Killing vector $\partial_t$ selects a preferred notion of positive frequency and allows for a Fock representation of the algebra. In this representation, the field operator admits the standard mode expansion
\begin{align}
    \hat{\phi}(t, \textbf{x}) = \int \frac{d^3 \textbf{k}}{(2\pi)^{3/2} \sqrt{2 \omega_\textbf{k}}}  \left( \hat{a}_{\textbf{k}} e^{i k_{\mu} x^{\mu}} + \hat{a}^{\dagger}_{\textbf{k}} e^{-i k_{\mu} x^{\mu}} \right), \label{eq:modedecomp}
\end{align}
where $k_{\mu} x^{\mu} = - \omega_\textbf{k} t + \textbf{k}\cdot \textbf{x}$, $ \omega_\textbf{k} = \sqrt{k^2 + m^2}$ with $k^2 = \textbf{k}\cdot \textbf{k}$, and the creation and annihilation operators satisfy
\begin{equation}
[\hat{a}_{\mathbf{k}}, \hat{a}^\dagger_{\mathbf{k}'}] = \delta^{(3)}(\mathbf{k}-\mathbf{k}')\hat{\mathbb{I}}.
\end{equation}
The Hamiltonian of the free scalar field can be defined with respect to the timelike Killing vector $\xi^a = (\partial_t)^a$ as
\begin{align}
\hat{H}_{\phi} \equiv \int_{\Sigma_t} d^3 \textbf{x} \ N^a \xi^{b} : T_{a b}(\hat{\phi},\hat{\phi}): = \int d^3 \textbf{k} \ \omega_\textbf{k} \hat{a}^{\dagger}_{\textbf{k}} \hat{a}_{\textbf{k}},  \label{Eq:freeHamil}
\end{align}
where $T_{ab}$ denotes the stress–energy tensor of the scalar field, $:T:$ represents normal ordering, and $\Sigma_t$ is a spacelike hypersurface orthogonal to $N^{a} = \xi^a/\sqrt{- \xi_b \xi^b}$.

Once the algebra of observables has been specified, the theory must be supplemented with the definition of quantum states. In the algebraic framework, a state is given by a complex-linear functional $\Omega: \mathcal{A}_{\phi}(\mathcal{M}) \to \mathbb{C}$, which assigns expectation values to observables and satisfies the usual normalization and positivity conditions, i.e., $\Omega(\mathbb{I}) = 1, \ \ \Omega(\hat{A}^\dagger \hat{A}) \geq 0$ for all $\hat{A} \in \mathcal{A}_{\phi}(\mathcal{M})$. In particular, all physical information about the field can be encoded in the corresponding $n$-point correlation functions. For instance, the two-point Wightman function associated with a state $\Omega$ is defined as
\begin{equation}
\mathcal{W}(\mathsf{x},\mathsf{x}') = \Omega(\hat{\phi}(\mathsf{x})\hat{\phi}(\mathsf{x}')).
\end{equation}
The correspondence with a Hilbert-space formulation is provided by the Gelfand–Naimark–Segal reconstruction theorem. Given an algebraic state $\Omega$, the GNS construction yields a Hilbert space $\mathcal H_\Omega$, a $*$-representation $\pi_\Omega$, and a cyclic vector $|\Omega\rangle$ such that $ \Omega(A)=\langle\Omega|\pi_\Omega(A)|\Omega\rangle.$ In representations in which a state is normal, it may additionally be represented by a density operator $\hat{\rho}$, such that $\Omega(\hat{A}) = \Tr (\hat{A} \hat{\rho}) = \langle \hat{A} \rangle_\rho$. A distinguished example of such a state is provided by the vacuum state arising from the Fock representation associated with the mode decomposition of the field. In this case, the vacuum $|0\rangle$ can be defined by $\hat{a}_{\mathbf{k}}|0\rangle = 0$ for all $\mathbf{k}$, and the corresponding bosonic Fock space is constructed in the usual way. The associated Wightman function then encodes the vacuum correlations of the quantum field.

More generally, we will consider states describing thermal equilibrium at nonzero temperature $1/\beta$. In the algebraic approach, these are characterized as KMS states, denoted by $\Omega_{\beta}$, which define quasifree states satisfying the KMS condition with respect to the time evolution generated by the Hamiltonian $\hat{H}_{\phi}$~\cite{Haag1996}. The KMS condition can be formulated in terms of the two-point Wightman function, $\mathcal{W}_{\beta}(t,t') = \Omega_{\beta}(\hat{\phi}(t,\textbf{x})\hat{\phi}(t',\textbf{x}') ) = \langle \hat{\phi}(t,\textbf{x})\hat{\phi}(t',\textbf{x}')\rangle_{\beta}$, through the relation $\mathcal{W}_{\beta}(t - i\beta, t') = \mathcal{W}_{\beta}(t',t)$. Whenever the state is invariant under time translations, the Wightman function depends only on the time difference, $\mathcal{W}_{\beta}(t,t') = \mathcal{W}_{\beta}(t - t' \equiv \Delta t)$. Under this assumption, the KMS condition reduces to $\mathcal{W}_{\beta}(\Delta t - i\beta) = \mathcal{W}_{\beta}(- \Delta t).$ In the field mode decomposition, we have
\begin{align}
 \mathcal{W}_{\beta}(t,t')  = & \int \frac{d^3 \textbf{k}}{(2\pi)^{3} 2 \omega_\textbf{k}} \Big((1+n_{\textbf{k}})e^{ik_\mu (x^{\mu}-x'^{\mu})} \nonumber \\ & +n_{\textbf{k}} e^{-ik_\mu (x^{\mu}-x'^{\mu} )} \Big), \label{eq:KMSw}
\end{align}
with $n_{\textbf{k}}= 1/(e^{\beta \omega_{\textbf{k}}} - 1)$. Throughout the text, we will occasionally also denote the thermal state by $\hat{\rho}_{\beta}$ when this notation is convenient for formal manipulations. In such cases, $\hat{\rho}_{\beta}$ should be understood simply as a symbolic representation of the KMS state $\Omega_{\beta}$, rather than, in general, as a trace-class density operator on a Hilbert space. In particular, expressions involving $\hat{\rho}_{\beta}$ are not meant to assume that a Gibbs density operator exists for the quantum field. In contrast, for a trace class Gibbs operator, in which the thermal state can be represented by a genuine Gibbs density operator, we reserve the notation $\hat{\rho}_R$.

The UDW particle detector provides an operational framework to probe quantum fields through localized interactions. In what follows, we consider a detector whose trajectory follows the flow of the timelike Killing vector field $\partial_t$ in Minkowski spacetime, so that the detector is static. The detector is modeled as a two-level system evolving along a worldline $\tau \mapsto \mathsf{x}(\tau)$ associated with an inertial static observer $\mathcal{O}$. In this case, the proper time of the detector coincides with the coordinate time, i.e., $\tau = t$. Its free Hamiltonian, expressed in terms of the proper time $\tau$, is given by $\hat{H}_d = \omega \hat{\sigma}_+ \hat{\sigma}_-$, where $\omega > 0$ denotes the energy gap and $\hat{\sigma}_{\pm}$ are the ladder operators of the Pauli algebra.

The interaction between the detector and the scalar field $\hat{\phi}(\mathsf{x})$ is assumed to be localized around the detector's trajectory and is described, in the interaction picture, by
\begin{align}
\hat{H}_I(\tau) & = \int_{\Sigma_\tau} d^3 \mathbf{x} \, \hat{h}_I(\mathsf{x}(\tau)) \label{eq:H_I}
\\  & = \lambda \int_{\Sigma_\tau} d^3 \mathbf{x} \, \Lambda(\mathsf{x}) \hat{m}(\tau) \otimes \hat{\phi}(\mathsf{x}(\tau)), \nonumber
\end{align}
where the corresponding monopole operator reads
\begin{align}
\hat{m}(\tau) = \hat{\sigma}_+ e^{i \omega \tau} + \hat{\sigma}_- e^{-i \omega \tau},
\end{align}
with $\lambda$ being a coupling constant and $\Lambda(\mathsf{x}) = \chi(\tau)\psi(\mathbf{x})$ defines the spacetime smearing of the interaction. Here, $\psi(\mathbf{x})$ encodes the spatial smearing of the detector around the curve $\mathsf{x}(\tau)$, while the temporal profile of the interaction is determined by the switching function $\chi(\tau)$ of compact support or sufficiently rapid decay. Furthermore, we take the spacetime smearing function $\Lambda(\mathsf{x})$ to be real and $\hat{\phi}(\mathsf{x}(\tau))$ should be understood as the pullback of the field in a neighborhood of the worldline $\mathsf{x}(\tau)$.

The dynamics of the coupled system is then governed by the unitary evolution operator
\begin{align}
\hat{U}_I = \hat{\mathcal{T}} \exp\left(-i \int d^4 \mathsf{x} \, \hat{h}_I(\mathsf{x}(\tau)) \right), \label{eq:U_I}
\end{align}
where $\hat{\mathcal{T}}$ denotes time ordering. This formulation ensures a covariant description of the interaction between the quantum field and a spatially smeared detector~\cite{Martinez20, Martinez21}.

\subsection{Ramsey scheme protocol}

The common operational approach to characterize heat in quantum systems is based on a two-point measurement scheme, where heat is associated with energy variations of the reservoir~\cite{Jarzynski2004, Goold2014}, which is briefly reviewed here. 

Let us consider a bipartite setup composed of a system of interest $S$ and a reservoir $R$, both with finite degrees of freedom, with total Hamiltonian $\hat{H} = \hat{H}_S + \hat{H}_R + \hat{H}_{\text{int}}$. The protocol starts by preparing the global state in a product form, where the reservoir is initially in a thermal state, $\hat{\rho}_0 = \hat{\rho}_S \otimes e^{-\beta \hat{H}_R}/Z_R,$ with $Z_R = \mathrm{Tr}[e^{-\beta \hat{H}_R}]$. An initial projective measurement of the reservoir energy is performed, projecting onto the eigenspaces of $\hat{H}_R$. Denoting the corresponding projectors by $\hat{\Pi}^0_n$, the outcome $E^0_n$ occurs with probability $p_n = \Tr(\hat{\rho}_0 \hat{\Pi}^0_n).$ After this measurement, the composite system evolves unitarily under $\hat{U}_{\tau,0}$, leading to the conditional state $\hat{\rho}_n(\tau) = (\hat{U}_{\tau,0} \hat{\Pi}^0_n  \hat{\rho}_0 \hat{\Pi}^0_n \hat{U}_{\tau,0}^\dagger)/p_n.$ A second projective measurement of the reservoir energy at time $\tau$ yields an outcome $E^\tau_m$, associated with the projector $\hat{\Pi}^\tau_m$, with conditional probability $p_{m|n} = \Tr(\hat{\Pi}^\tau_m  \hat{\rho}_n(\tau)).$ Within this framework, heat is defined as the stochastic energy change in the reservoir, $Q_{m,n} = E^\tau_m - E^0_n.$ The corresponding probability distribution for the forward process is then given by
\begin{equation}
P(Q) = \sum_{m,n} p_{m,n} \delta\left(Q - Q_{m,n}\right),
\end{equation}
where $p_{m,n} = p_n p_{m|n}$ denotes the joint probability of observing outcomes $(m,n)$.
As a result, the stochastic variable $Q$ captures the energy exchanged with the environment, providing an operational notion of heat consistent with quantum thermodynamics. In particular, the probability distribution $P(Q)$ fully characterizes the heat statistics and, as expected, its first moment yields the average dissipated heat. Indeed, one finds
\begin{align}
    \langle Q \rangle = \int dQ P(Q) Q =\mathrm{Tr}[\hat{H}_R(\hat{\rho}'_R - \hat{\rho}_R)]
\end{align}
where $\hat{H}_R$ is time-independent and $\hat{\rho}'_R = \Tr_S\left(\hat{U}_{\tau,0}\hat{\rho}_0 \hat{U}_{\tau,0}^\dagger \right)$ denotes the reduced state of the reservoir after the interaction. Moreover, the corresponding characteristic function is defined as
\begin{align}
    C(\nu) = \int dQ P(Q) e^{i\nu Q},
\end{align}
which, for the protocol described above, can be written as
\begin{align}
    C(\nu) = \Tr \left( \hat{U}_{\tau,0}^\dagger e^{i\nu\hat{H}_R} \hat{U}_{\tau,0} e^{-i \nu \hat{H}_R}(\hat{\rho}_S \otimes \hat{\rho}_R)\right).
\end{align}

Despite its operational appeal, the TPM scheme presents some limitations. In the context of relativistic quantum field theory, its direct implementation is problematic due to its reliance on instantaneous projective measurements, which generically lead to violations of relativistic causality, as discussed in the introduction. In particular, projective measurements on quantum fields can induce superluminal signalling, rendering the TPM protocol incompatible with a fully relativistic framework, as already noted by Ortega \textit{et al.}~\cite{Ortega} in the context of work distributions. Moreover, even in the case of reservoirs with finite degrees of freedom, the implementation of two projective energy measurements can be experimentally demanding or impractical, especially when dealing with large or complex environments. These limitations motivate the search for alternative formulations that avoid explicit projective measurements while still providing access to the full statistics of heat exchange. A key observation is that the full information about the heat distribution is encoded in its characteristic function. This suggests an alternative strategy in which, rather than performing direct projective measurements, one reconstructs the characteristic function through an interferometric protocol. In this spirit, Ramsey interferometry provides a powerful framework to access thermodynamic quantities in a fully unitary and coherent manner~\cite{Dorner2013, Mazzola2013, Goold2014}. In particular, it replaces intermediate projective measurements with controlled unitary operations and quantum coherence, thereby preserving the causal structure of the theory. 

Motivated by these developments, we now adopt an interferometric approach to define the heat distribution for a quantum field interacting with a localized detector by following Ref.~\cite{Goold2014}, which introduced a Ramsey-based protocol to define heat distributions for quantum systems with finite degrees of freedom. Rather than describing the protocol from a finite-dimensional setting, we directly formulate it in the quantum field theoretic framework using the notation introduced in the previous section. This allows us to consistently describe heat exchange in a relativistic setting while maintaining an operational interpretation in terms of localized measurements.

\begin{figure}[t]
    \centering
    \includegraphics[width=0.75\linewidth]{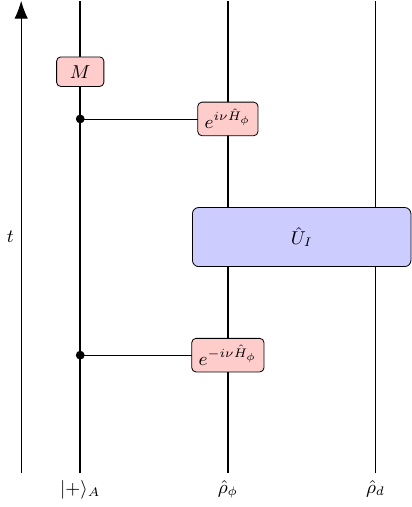}
    \caption{Schematic representation of the interferometric protocol used to access the characteristic function of heat. An ancilla prepared in the state $\ket{+}_A$ controls the operations $\hat{V}(\nu)$ and $\hat{V}^\dagger(\nu)$ acting on the quantum field, while the unitary $\hat{U}_I$ describes the interaction between the field and the UDW detector. The heat statistics is encoded in the final measurement performed on the ancilla.}
    \label{fig:heatprotocol}
\end{figure}

The central idea of the interferometric protocol is to encode the characteristic function of the heat distribution, $C(\nu)$, into the coherence of an auxiliary quantum system. In the present setting, this auxiliary system $A$ can be modeled by a second UDW detector, which plays the role of an ancilla. 
The system of interest is another UDW detector $D$, initially prepared in an arbitrary state $ \hat{\rho}_{d,0}$, while the reservoir is given by the quantum field $\hat{\phi}(\mathsf{x})$, initially prepared in a thermal KMS state, symbolic represented by $\hat{\rho}_\beta$.

The protocol is represented diagrammatically in Fig.~\ref{fig:heatprotocol} and proceeds as follows. The ancilla detector $A$ is initially prepared in the superposition state $|+\rangle_A = \frac{1}{\sqrt{2}}\left(|0\rangle_A + |1\rangle_A\right),$ which can be generated by applying a Hadamard gate to its ground state. The next step consists of a controlled unitary operation between the ancilla and the field, implementing the transformation
\begin{align}
\hat{V}(\nu) = |0\rangle_A\langle 0| \otimes \hat{\mathbb{I}}_{\phi} + |1\rangle_A\langle 1| \otimes e^{-i\nu \hat{H}_\phi},
\end{align}
where $\hat{H}_\phi$ is the free Hamiltonian of the field given by Eq.~\eqref{Eq:freeHamil}. Next, the field interacts with the UDW detector $D$ through the unitary evolution operator $\hat{U}_I$ given by Eq.~\eqref{eq:U_I}, which describes the localized interaction between the detector and the quantum field. This step encodes the exchange of energy between the detector and the field. After this interaction, a second controlled operation between the ancilla and the field is applied,
\begin{align}
\hat{V}^{\dagger}(\nu) = |0\rangle_A\langle 0| \otimes \hat{\mathbb{I}}_{\phi} + |1\rangle_A\langle 1| \otimes e^{i\nu \hat{H}_\phi}.
\end{align}
thereby completing the interferometric sequence. At the end of the protocol, the tripartite state of ancilla, detector, and field is given by
\begin{align}
    \hat{\rho}_{A,d,\phi} =\hat{V}^{\dagger} \hat{U}_I \hat{V} \left(|+\rangle_A \langle + | \otimes \hat{\rho}_{d,0} \otimes  \hat{\rho}_\beta \right) \hat{V}^{\dagger} \hat{U}^{\dagger}_I \hat{V} 
\end{align}
Tracing out the detector and field degrees of freedom, the off-diagonal elements of the reduced ancilla state encode the characteristic function of the heat distribution.

As introduced in Sec.~\ref{sec:setting}, let $\mathcal A_\phi$ denote the algebra of field observables and $\Omega_\beta$ the corresponding thermal KMS state. We further denote by $\tau_\phi^\nu$ the free time-translation automorphism associated with the field Hamiltonian introduced in Eq.~\eqref{Eq:freeHamil}, with respect to which $\Omega_\beta$ is a $(\tau_\phi,\beta)$-KMS state. On
the joint algebra $\mathcal O =\mathcal B(\mathcal H_d)\otimes\mathcal A_\phi$, where $\mathcal B(\mathcal H_d)$ is the algebra of bounded operators acting on the detector Hilbert space, the initial state is written as $\Omega_0=\rho_{d,0}\otimes\Omega_\beta,$ where $\rho_{d,0}$ denotes the detector state in the algebraic formulation, corresponding to the density operator $\hat{\rho}_{d,0}$ acting on $\mathcal H_d$. The characteristic function can be
defined algebraically by
\begin{equation}
  C(\nu) =   \Omega_0\!\left[\hat U_I^\dagger\bigl(\mathrm{id}_d \otimes\tau_\phi^\nu\bigr)(\hat U_I)
  \right].
  \label{eq:C(nu)algebraic}
\end{equation}
Using the symbolic density-operator notation $\hat{\rho}_\beta$ for the field KMS state $\Omega_\beta$, the algebraic expression above can be formally represented as
\begin{align}
C(\nu) = \Tr \left(\hat{U}_I^\dagger e^{i\nu \hat{H}_{\phi}} \hat{U}_I e^{-i \nu \hat{H}_{\phi}}(\hat{\rho}_{d,0} \otimes \hat{\rho}_{\beta})\right). \label{Eq:C(nu)}
\end{align}
This is the trace notation used in the perturbative calculations below. As emphasized above, $\hat{\rho}_\beta$ is only a symbolic representation of the field KMS state and is not assumed to define a trace-class density operator.

%In a confined-field regularization, in which
%\(\omega_\beta\) is represented by
%\(\hat\rho_\beta=e^{-\beta\hat H_\phi}/Z_\phi\), this leads to the expression
%\begin{align}
%C(\nu) = \Tr \left(\hat{U}_I^\dagger e^{i\nu \hat{H}_{\phi}} \hat{U}_I e^{-i \nu \hat{H}_{\phi}}(\hat{\rho}_{d,0} \otimes \hat{\rho}_{\beta})\right). \label{Eq:C(nu)}
%\end{align} which is the trace representation used in the perturbative calculations
%below.  Thus \(\hat\rho_\beta\) and \(Z_\phi\) are not assumed to exist as
%trace-class objects in the infinite-volume theory. }

Following the same philosophy adopted by Ref.~\cite{Ortega} for the work distribution, once the characteristic function of the field is well defined independently of the TPM scheme, the heat distribution in the Ramsey protocol can be defined as the inverse Fourier transform of $C(\nu)$, i.e., 
\begin{align}
P(Q) = \frac{1}{2 \pi}\int_{-\infty}^{\infty} d\nu  C(\nu) e^{-i \nu Q}. \label{Eq:P(Q)}
\end{align}
In this way, the full statistics of heat exchange are reconstructed from the interferometric protocol, providing a consistent and operational notion of heat for quantum fields probed by localized detectors. Moreover, as expected, the average heat exchanged can be obtained from the first moment (or cumulant) of the characteristic function, yielding the standard thermodynamic definition, i.e., 
\begin{align}
\langle Q \rangle & \equiv i^{-1} \left(\frac{d C(\nu)}{d \nu}\right)_{\nu = 0} \label{Eq:aveQ}\\ & = i^{-1} \left(\frac{d}{d \nu} \mathrm{Tr} \left[ \hat{U}_I^\dagger e^{i\nu \hat{H}_{\phi}} \hat{U}_I e^{-i \nu \hat{H}_{\phi}}(\hat{\rho}_{d,0} \otimes \hat{\rho}_{\beta})\right] \right)_{\nu = 0} \nonumber\\
& = \mathrm{Tr} \left[ (\hat{U}_I^\dagger \hat{H}_{\phi} \hat{U}_I - \hat{U}_I^\dagger \hat{U}_I \hat{H}_{\phi})(\hat{\rho}_{d,0} \otimes \hat{\rho}_{\beta})\right]\nonumber \\
& = \mathrm{Tr}_{\phi}[\hat{H}_{\phi}(\hat{\rho}_{\phi,\tau}  -  \hat{\rho}_{\beta} )]. \nonumber
\end{align}
where $\hat{\rho}_{\phi, \tau} = \Tr_d\left(\hat{U}_I \hat{\rho}_{d,0} \otimes \hat{\rho}_{\beta} \hat{U}_I^\dagger\right)$.

This is the quantity denoted by $\Delta \mathcal{Q}$ in
Ref.~\cite{JaksicPillet2014}, specialized here to the quantum
field as the thermal reservoir. Indeed, denoting by $\delta_\phi$ the generator of the free-field dynamics, $\tau_\phi^\nu=e^{\nu\delta_\phi}$, Eqs.~\eqref{eq:C(nu)algebraic} and~\eqref{Eq:aveQ} allow us to obtain
\begin{equation}
    \langle Q\rangle
    =
    -i\,\Omega_0\!\left[
    \hat U_I^\dagger
    (\mathrm{id}_d\otimes\delta_\phi)(\hat U_I)
    \right].
\end{equation}
Upon identifying $\delta_R=\mathrm{id}_d\otimes\delta_\phi$, this is precisely the algebraic definition
$\Delta\mathcal{Q}=-i\,\Omega_i(U^*\delta_R(U))$ of
Ref.~\cite{JaksicPillet2014}. In a Hamiltonian representation,
$\delta_\phi(A)=i[\hat H_\phi,A]$, this reduces to the energy variation in Eq.~\eqref{Eq:aveQ}.

\subsection{Algebraic entropy balance and Landauer bounds}
\label{sec:landauer-bounds}

The interferometric protocol introduced above can also be employed to derive thermodynamic constraints on the heat exchanged during information-processing tasks. In particular, it provides a natural framework to investigate Landauer-type bounds directly from the statistics of heat exchange, in the spirit of fluctuation relations~\cite{Goold2015}. A rigorous formulation of Landauer's principle was given by Reeb and Wolf~\cite{Reeb2014}. In their setting, a finite-dimensional system interacts unitarily with a finite-dimensional reservoir initially prepared in a Gibbs state. The resulting entropy balance relates the heat absorbed by the reservoir to the entropy change of the system, together with contributions accounting for correlations generated during the interaction and for the departure of the final reservoir state from thermal equilibrium. The usual Landauer bound then follows from the positivity of these contributions. 

The quantum-field setting considered here requires a different formulation, since the reservoir is an infinitely extended field in a KMS state and, in general, cannot be described by a trace-class Gibbs density operator. We therefore employ the algebraic entropy-balance framework of Jak\v{s}i\'c and Pillet~\cite{JaksicPillet2014}. This framework applies precisely to our setting, in which the detector is a finite-dimensional system while the quantum field plays the role of an infinitely extended reservoir initially prepared in a KMS state, once we associate the $*$-algebra introduced in Sec.~\ref{sec:setting} with the corresponding $C^*$-algebra $\mathcal A_\phi^{\rm W}$, generated by the exponentiated smeared field operators, and define the joint algebra as $\mathcal O_{\rm W}=\mathcal B(\mathcal H_d)\otimes\mathcal A_\phi^{\rm W}$.

Thus, let us consider the final joint state and its detector restriction
\begin{equation}
  \Omega_\tau(O) = \Omega_0\!\left(\hat U_I^\dagger O\hat U_I\right), \qquad   \rho_{d,\tau} =   \left.\Omega_\tau\right|_{\mathcal B(\mathcal H_d)},
\end{equation}
for $O \in \mathcal{O}_W$ and use the entropy-change convention
$ \Delta S = S(\rho_{d,\tau})-S(\rho_{d,0}).$  To apply the entropy-balance theorem of Jakšić and Pillet, we assume that $\hat{U}_I\in\mathcal U(\mathcal O_{\rm W})\cap\operatorname{Dom}\,(\mathrm{id}_d \otimes \delta_\phi),$ where $\mathcal U(\mathcal O_{\rm W})$ denotes the group of unitary elements of the joint $\mathcal O_{\rm W}$, $\delta_\phi$ is the infinitesimal generator of the free-field dynamics $\tau_\phi^t$, and $\operatorname{Dom}\,(\mathrm{id}_d \otimes \delta_\phi)$ is the domain of the generator of the free-field dynamics, extended trivially to the detector sector. Then, the theorem gives
\begin{equation}
  \Sigma = D_{\mathrm{Ar}}\!\left(\Omega_\tau\,\middle\|\,
    \rho_{d,\tau}\otimes\Omega_\beta\right)
  = \beta\langle Q\rangle+\Delta S.
\end{equation}
Here $D_{\mathrm{Ar}}$ denotes the Araki relative entropy~\cite{Araki1976,Araki1977}, which provides the algebraic generalization of quantum relative entropy to states on operator algebras and is defined through the relative modular operator associated with the pair of states (see Ref.~\cite{Witten2018} for a recent review). Its non-negativity yields finally
\begin{equation}
  \beta\langle Q\rangle\geq-\Delta S 
  \label{eq:Landauer}
\end{equation}
as a direct application of the algebraic theorem. No ergodicity assumption is needed for the entropy-balance identity or inequality. 

For comparison, consider a regularized type-I reservoir for which the thermal KMS state is represented by a trace-class Gibbs operator $\hat\rho_\beta \to \hat{\rho}_R = e^{-\beta\hat H_R}/Z_R$~\cite{Haag1996}. In what follows, hats are used explicitly to emphasize that, in this regularized setting, the states are genuine density operators on a Hilbert space. Accordingly, the Araki relative entropy $D_{\mathrm{Ar}}$ reduces to the usual quantum relative entropy between density operators $D$. Given $\hat\rho_\tau = \hat U_I \bigl(\hat\rho_{d,0}\otimes\hat\rho_R\bigr) \hat U_I^\dagger,$
a direct calculation gives
\begin{align}
  D\!\left(
    \hat\rho_\tau\,\middle\|\,
    \hat\rho_{d,\tau}\otimes\hat\rho_R
  \right)  &= - S(\hat\rho_\tau)+S(\hat\rho_{d,\tau}) - \Tr\bigl(\hat\rho_{\phi,\tau}\ln\hat\rho_R\bigr)\nonumber \\ 
  &=   \Delta S+\beta\langle Q\rangle.
\end{align}
In the same regularization, the entropy production decomposes as
\begin{equation}
  \Sigma
  =  I(\hat{\rho}_{d,\tau}:\hat{\rho}_{\phi,\tau}) + D\!\left(\hat\rho_{\phi,\tau}\middle\|\hat\rho_R\right),
\end{equation}
which is the Reeb--Wolf identity~\cite{Reeb2014}, with $I(\hat{\rho}_{d,\tau}:\hat{\rho}_{\phi,\tau})$ being the quantum mutual information between the detector and the field and $D\!\left(\hat\rho_{\phi,\tau}\middle\|\hat\rho_R\right)$ representing the departure of the final reservoir state from thermal equilibrium.

There is a second lower bound that follows from the field-energy
full-counting statistics~\cite{Goold2015}. If $P(Q)$ is normalized and has a finite exponential moment at $\beta$, then
\begin{equation}
  \left\langle e^{-\beta Q}\right\rangle
  =   \int_{-\infty}^{\infty}dQ\,P(Q)e^{-\beta Q}
  =   C(i\beta),
  \label{eq:flucrel}
\end{equation}
where $C(i\beta)$ denotes the analytic continuation of $C(\nu)$ from real $\nu$.    Jensen's inequality then gives
\begin{equation}
   \beta\langle Q\rangle\geq-\ln C(i\beta). 
  \label{eq:Landauer1}
\end{equation}
Equation~\eqref{eq:Landauer} follows from the algebraic
entropy-balance theorem, whereas Eq.~\eqref{eq:Landauer1} follows from the normalized heat distribution and Jensen's inequality. %Alternatively, the former may be established in a finite-dimensional regularization and subsequently extended to the field limit, provided that the mean heat and the reduced detector state converge as the regulators are removed. 

In Sec.~\ref{sec:IIIB}, we evaluate the bounds given by Eqs.~\eqref{eq:Landauer} and~\eqref{eq:Landauer1} for the finite-temperature detector-field interaction model using the mean field-energy gain and the detector reduced dynamics computed up to second-order in perturbation theory. This provides a model-specific perturbative verification of the bounds and allows us to compare their tightness for different detector preparations and interaction times.  The smooth switching and smearing profiles employed ensure that all perturbative integrals are finite, although this finiteness alone does not establish the algebra-membership and domain conditions assumed above for the exact interaction unitary. Although $\hat U_I$ is a bounded operator, showing that it belongs to the domain of the generator of the free-field dynamics is nontrivial and is not addressed here. Here, it is taken as an assumption. A potentially rigorous setting is the $W^*$-dynamical extension for unbounded perturbations mentioned in Remark 1 of Sec.~4.2 of Ref.~\cite{JaksicPillet2014} and developed in Ref.~\cite{Dere2003}, which may provide a more natural setting for the UDW interaction Hamiltonian. A detailed investigation of its application to the explicitly switched interaction is left for future work.

\subsection{The laws of quantum thermodynamics for detector--field dynamics}
Because the UDW interaction is explicitly switched in time, the energy gained by the field $\langle Q\rangle$ should not, in general, be identified with the energy lost by the detector $-\Delta E_d$. 

Indeed, considering the total Hamiltonian $\hat H(\tau)=\hat H_d+\hat H_\phi+\hat H_I(\tau),$ the explicit time dependence of $\hat H_I(\tau)$ implies that an external agent responsible for controlling the switching can perform work on the detector--field system. For a unitary evolution, the corresponding mean switching work is
$$ \langle W_{\mathrm{sw}}\rangle = \int_{\tau_0}^{\tau_1} d\tau\, \left\langle \frac{\partial \hat H(\tau)}{\partial \tau} \right\rangle = \int_{\tau_0}^{\tau_1}\, \left\langle \frac{\partial \hat H_I(\tau)}{\partial \tau} \right\rangle . $$
If the interaction Hamiltonian vanishes at the initial and final times, $\hat H_I(\tau_0)=\hat H_I(\tau_1)=0$, the change in the total energy reduces to the sum of the changes in the free detector and field energies. Since $\langle Q \rangle$ is defined here as the energy gained by the field, one obtains
\begin{align}
\langle W_{\mathrm{sw}}\rangle =
\Delta E_d+\langle Q\rangle , \label{eq:firstlaw}
\end{align}
where $\Delta E_d = \operatorname{Tr}\left(\hat H_d(\hat\rho_{d,\tau}-\hat\rho_{d,0})\right)$ is the variation of the internal energy of the detector. This relation may be regarded as the first-law energy balance for the externally driven detector--field system. Equivalently, defining $Q_d=-Q$ as the heat absorbed by the detector, it takes the conventional form of the first law, i.e., $\Delta E_d=\langle W_{\mathrm{sw}}\rangle+Q_d.$ Thus, at finite interaction times, part or even all of the energy deposited in the field may originate from the external switching agent rather than from a decrease in the detector energy.

Moreover, using this energy balance, the Landauer relation in Eq.~\eqref{eq:Landauer}, can be equivalently written as

\begin{align}
  \beta\bigl(\langle W_{\mathrm{sw}}\rangle-\Delta F_d
  \bigr)\geq0,  \label{eq:secondlaw} 
\end{align}
where $\Delta F_d = \Delta E_d-\beta^{-1}\Delta S$ is the change in the nonequilibrium free-energy functional of the detector~, evaluated at the reservoir temperature~\cite{Esposito2010, Chaves2025}. Hence, $\langle W_{\mathrm{sw}}\rangle\geq\Delta F_d,$ so that the entropy production can be identified with the dissipated switching work, $\Sigma=\beta W_{\mathrm{diss}}$, where $W_{\mathrm{diss}} = \langle W_{\mathrm{sw}}\rangle-\Delta F_d$. In this sense, while the previous energy-balance relation~\eqref{eq:firstlaw} provides the first-law statement for the finite-time UDW protocol, the inequality~\eqref{eq:secondlaw} gives its corresponding nonequilibrium second-law formulation. 

This interpretation is consistent with the finite-time UDW analysis of Ref.~\cite{Shevchenko2017}, where the same underlying energy balance and free-energy bound are discussed, although they are not explicitly formulated there as first- and second-law statements for the detector--field dynamics. Throughout this work, we retain the conventional term \emph{heat} for the reservoir energy gain $\langle Q \rangle$, while keeping explicit that, for a finite-time interaction, the external switching agent can supply part or all of this energy.

\section{Perturbative Evaluation of Heat Statistics}
\label{Sec:III}
This section is divided into two main subsections. We begin by analyzing the case in which the field is prepared in the vacuum state, and subsequently consider thermal (KMS) states at finite temperature. Although the vacuum can be formally recovered as a limiting case of a KMS state in the zero-temperature limit, it is useful to treat it separately. In particular, the vacuum case provides a simpler setting that allows us to clearly establish conventions for the sign of heat exchange, thereby fixing what is meant by positive and negative dissipated heat.

\subsection{Vacuum $\beta \to \infty$}

We begin by considering the case in which the quantum field is initially prepared in the vacuum state $\hat{\rho}_{\text{vac}} = |0\rangle \langle0|$. In this setting, we compute the characteristic function of the heat distribution and the corresponding probability distribution up to second order in the coupling constant $\mathcal{O}(\lambda^2)$. This perturbative analysis provides the leading-order contribution to the heat exchange between the detector and the field, and establishes the basic structure that will later be extended to thermal KMS states. The main steps in the evaluation of $C(\nu)$ up to $\mathcal{O}(\lambda^2)$ are as follows. They closely follow the derivation presented in the Appendix of Ref.~\cite{Ortega} for the work distribution. However, the present treatment is more general, as the detector is not assumed to be gapless, and the full time-dependent structure of the monopole operator $\hat{m}(\tau)$ must be explicitly taken into account.

Let us begin by expanding the time-evolution operator given by Eq.~\eqref{eq:U_I} in a Dyson series
\begin{align}
    \hat{U}_I = \hat{\mathbb{I}} + \hat{U}^{(1)}_I + \hat{U}^{(2)}_I + \mathcal{O}(\lambda^3),
\end{align}
with
\begin{align}
     & \hat{U}^{(1)}_I = -i\int d^4 \mathsf{x} \, \hat{h}_I(\mathsf{x}(\tau)), \\
     &\hat{U}^{(2)}_I =  - \iint d^4 \mathsf{x} \, d^4 \mathsf{x}' \, \hat{h}_I(\mathsf{x}(\tau)) \hat{h}_I(\mathsf{x}(\tau')) \Theta(\tau - \tau'), \nonumber
\end{align}
where $\Theta(\tau - \tau')$ is the Heaviside step function. Substituting this expansion into Eq.~\eqref{Eq:C(nu)}, we obtain
\begin{align}
    C(\nu) = & 1 + \mathrm{Tr}\left(\hat{U}^{(2)}_I  \hat{\rho}_{d,0} \otimes \hat{\rho}_{\text{vac}}\right) + \mathrm{Tr}\left(\hat{U}^{(2) \dagger}_I  \hat{\rho}_{d,0} \otimes \hat{\rho}_{\text{vac}} \right) \nonumber \\ &+ \mathrm{Tr}\left(\hat{U}_I^{(1)\dagger} e^{i\nu \hat{H}_{\phi}} \hat{U}^{(1)}_I e^{-i \nu \hat{H}_{\phi}}(\hat{\rho}_{d,0} \otimes \hat{\rho}_{\text{vac}})\right), \label{eq:Cdyson}
\end{align}
by noticing that all terms at $\mathcal{O}(\lambda)$ vanish, since the vacuum is a quasifree state and, in particular, $\langle 0| \hat{\phi}(\textsf{x})|0\rangle = 0$.

Now, the terms in the right-hand side of Eq.~\eqref{eq:Cdyson} can be written as
\begin{widetext}
{\small
\begin{align}
     & \mathrm{Tr}\left(\hat{U}_I^{(1)\dagger} e^{i\nu \hat{H}_{\phi}} \hat{U}^{(1)}_I e^{-i \nu \hat{H}_{\phi}}(\hat{\rho}_{d,0} \otimes \hat{\rho}_{\text{vac}})\right) = \lambda^2 \iint d^4 \mathsf{x} \, d^4 \mathsf{x}' \Lambda(\mathsf{x})\Lambda(\mathsf{x}')  \Tr_d\left(\hat{m}(\tau)\hat{m}(\tau') \hat{\rho}_{d,0}  \right) \mathcal{W}_{\text{vac}}(\tau, \tau'+\nu), \\
     &     \mathrm{Tr}\left(\hat{U}^{(2)}_I  \hat{\rho}_{d,0} \otimes \hat{\rho}_{\text{vac}}\right)  + \mathrm{Tr}\left(\hat{U}^{(2) \dagger}_I  \hat{\rho}_{d,0} \otimes \hat{\rho}_{\text{vac}} \right) =  - 2 \lambda^2 \mathfrak{R}  \left[\iint d^4 \mathsf{x} \, d^4 \mathsf{x}' \Lambda(\mathsf{x}) \Lambda(\mathsf{x}') \Theta(\tau - \tau')\mathrm{Tr}_d\left(\hat{m}(\tau)\hat{m}(\tau') \hat{\rho}_{d,0}  \right) \mathcal{W}_{\text{vac}}(\tau, \tau')\right],     \nonumber
\end{align}
}
where $\mathcal{W}_{\text{vac}}(\tau, \tau'+\nu) =  \langle\hat{\phi}(\tau,\textbf{x}) \hat{\phi}(\tau' + \nu,\textbf{x})\rangle_{\text{vac}}$ and $\mathfrak{R}$ denotes the real part of the argument. This allows us to write $C(\nu)$ as
\begin{align}
    C(\nu) =  1 + \lambda^2 \iint d^4 \mathsf{x} \, d^4 \mathsf{x}' \Lambda(\mathsf{x}) \Lambda(\mathsf{x}')
    \Big\{ & \mathrm{Tr}_d\left(\hat{m}(\tau)\hat{m}(\tau') \hat{\rho}_{d,0}  \right) \mathcal{W}_{\text{vac}}(\tau, \tau'+\nu) \label{Eq:Cgeneral}\\& - 2 \mathfrak{R}\left[\Theta(\tau - \tau') \mathrm{Tr}_d\left(\hat{m}(\tau)\hat{m}(\tau') \hat{\rho}_{d,0}  \right) \mathcal{W}_{\text{vac}}(\tau, \tau')\right]\Big\}. \nonumber
\end{align} 
\end{widetext}

Finally, using the fact that
\begin{align}
  \Tr_d\left(\hat{m}(\tau)\hat{m}(\tau') \hat{\rho}_{d,0}  \right) = e^{i \omega(\tau-\tau')}  \rho_{11,d} + e^{-i \omega(\tau-\tau')}  \rho_{00,d},   \nonumber
\end{align}
where $\rho_{jj,d} = \langle j | \hat{\rho}_{d,0} | j \rangle, \ j = 0,1$ denote the diagonal elements of the detector's initial density matrix, corresponding to the probabilities that the detector is initially in the ground or excited state, together with the Fourier representation of $\Theta(\tau - \tau' = u)$, i.e., 
\begin{align}
    \Theta(u) = \frac{1}{2 \pi} \int_{-\infty}^{\infty}\left(\pi  \delta(\Omega) - i \mathcal{P}\left(\frac{1}{\Omega} \right) \right) e^{-i \Omega u}d\Omega,
\end{align}
where $\mathcal{P}$ denotes Cauchy principal value, and Eq.~\eqref{eq:KMSw} in the limit of $\beta \to \infty$, the characteristic function at $\mathcal{O}(\lambda^2)$ can be expressed as
\begin{widetext}
\begin{align}
    C(\nu)    & = 1 + \lambda^2 \int \frac{d^3 \textbf{k}}{(2\pi)^3 2 \omega_{\textbf{k}}}\left(|\tilde{\Lambda}(\omega - \omega_{\textbf{k}}, \textbf{k})|^2\rho_{11,d} + |\tilde{\Lambda}(\omega + \omega_{\textbf{k}}, \textbf{k})|^2\rho_{00,d}\right) (e^{i \nu \omega_{\textbf{k}}} - 1)
 \end{align}
\end{widetext}
where
\begin{align}
    \tilde{\Lambda}(\omega \pm \omega_{\textbf{k}}, \textbf{k}) = \tilde{\chi}(\omega \pm \omega_{\textbf{k}}) \tilde{\psi}(\textbf{k}),
\end{align}
with $\tilde{\psi}(\textbf{k})$ and $\tilde{\chi}(\omega \pm \omega_{\textbf{k}})$ being the Fourier transforms of $\psi(\textbf{x)}$ and $\chi(\tau)$, respectively. One can see that $C(\nu)$ is independent of $\beta$ in the limit of zero temperature. In this case, it is worth noting that the term $|\tilde{\chi}(\omega - \omega_{\mathbf{k}})|^2 \rho_{11,d}$ in the integrand corresponds to a detector with energy gap $\omega$ emitting a wavepacket whose momentum components are centred around the resonance condition $\omega \approx \omega_{\mathbf{k}}$, thereby exciting the field (de-excitation of the detector). On the other hand, the term $|\tilde{\chi}(\omega + \omega_{\mathbf{k}})|^2 \rho_{00,d}$ does not exhibit resonances when $\omega > 0$.  This reflects the fact that the probability of exciting an inertial detector through interaction with the vacuum is significantly smaller than the probability of its de-excitation. Otherwise, it would correspond to the simultaneous excitation of the detector and the field.  Moreover, it is worth noting that $C(\nu)$ resembles the characteristic function of a Poisson distribution for a stochastic process up to order $\mathcal{O}(\lambda^2)$~\cite{Hida2008, Esposito2009}. 

By taking the inverse Fourier transform of $C(\nu)$, we obtain the probability distribution of heat
\begin{align}
P(Q) =  P(Q=0) \delta(Q) + P(Q>0)  \delta(Q-\omega_{\mathbf{k}}),
\end{align}
where
\begin{align}
& P(Q=0) = 1 - P(Q>0), \nonumber \\
&   P(Q>0)  = \lambda^2 \int \frac{d^3 \mathbf{k}}{(2\pi)^3} \frac{|\tilde{\psi}(\mathbf{k})|^2}{2\omega_{\mathbf{k}}} \mathcal{F}(\omega,\omega_{\mathbf{k}}),\\ & \mathcal{F}(\omega,\omega_{\mathbf{k}}) = |\tilde{\chi}(\omega - \omega_{\textbf{k}})|^2\rho_{11,d} + |\tilde{\chi}(\omega + \omega_{\textbf{k}})|^2\rho_{00,d}, \nonumber
\end{align}
Let us notice that $P(Q>0)$ represents the probability of energy (heat) exchange between the detector and the field. This contribution arises mainly from the detector deexcitation term $|\tilde{\chi}(\omega - \omega_{\mathbf{k}})|^2 \rho_{11,d}$ and from the counter-resonant term $|\tilde{\chi}(\omega + \omega_{\mathbf{k}})|^2 \rho_{00,d}$. In contrast, $P(Q=0) = 1 - P(Q>0)$ represents the probability that no heat exchange occurs between the detector and the field. Hence, the heat distribution is composed of a peak at $Q=0$ with probability $P(Q=0)$, corresponding to no heat exchange, and a positive continuum with a peak at $Q=\omega_{\mathbf{k}}$, with probability $P(Q>0)$, corresponding to processes in which energy is transferred from the detector to the field. In particular, within this convention, the dissipated heat is positive when the detector undergoes a deexcitation and the field absorbs a quantum of energy. Moreover, one can easily verify that $\int_{-\infty}^{\infty} dQ \, P(Q) = 1$. Although we refer to $P(Q)$ as a probability throughout the manuscript, it is in fact a probability density with dimensions of inverse heat (or inverse energy), such that $P(Q)dQ$ is dimensionless.

The average heat exchange can be calculated from the first moment (or cumulant) of the characteristic function in Eq.~\eqref{Eq:aveQ} yielding
\begin{align}
   \langle Q \rangle = \frac{\lambda^2}{2} \int \frac{d^3 \mathbf{k}}{(2\pi)^3}  |\tilde{\psi}(\mathbf{k})|^2 \mathcal{F}(\omega, \omega_{\textbf{k}}), \label{eq:aveQvac}
\end{align}
which implies that $\langle Q \rangle \ge 0$.

\subsubsection{Massless scalar field}

We now specialize to the case of a massless scalar field, for which  $\omega_{\mathbf{k}} = ||\textbf{k}|| = k$, and adopt Gaussian functions for the spacetime smearing. This choice allows for analytic control over the integrals involved, while providing a physically well-motivated description of a localized detector with smooth switching and spatial profile. In particular, the Gaussian smearing ensures ultraviolet regularity and simplifies the evaluation of the heat distribution in the perturbative regime. Let
\begin{align}
    & \chi(t) = e^{-t^2/2T^2}, \ \ \ \lim_{T \to \infty} \chi(t) = 1, \label{eq:gauss_switching} \\
    & \psi(\textbf{x}) = \frac{e^{-|\textbf{x}|^2/2 \sigma^2}}{\sqrt{(2\pi \sigma^2)^3}},  \ \ \ \ \lim_{\sigma \to 0}  \psi(\textbf{x}) = \delta(\textbf{x}). \label{eq:gauss_smearing}
\end{align}
such that the Fourier transform of the spacetime smearing functions are given by
\begin{align}
    & \tilde{\chi}(\Omega) = \sqrt{2\pi} T e^{-T^2 \Omega^2/2},\\
    & \tilde{\psi}(\textbf{k}) = e^{-\sigma^2|\textbf{k}|^2/2}
\end{align}

From these choices, a straight forward calculation shows that the probability of positive heat contribution $P_+(Q) = P(Q>0)\,\delta(Q-k)$ can be written as
\begin{align}
    P_+(Q) & = \frac{\lambda^2}{4 \pi^2} \int_0^{\infty} dk k e^{- \sigma^2 k^2} \mathcal{F}(\omega, k) \delta(Q-k) \nonumber \\ & =  \frac{\lambda^2}{4 \pi^2}  Q e^{- \sigma^2 Q^2} \mathcal{F}(\omega, Q),
\end{align}
with
\begin{align}
    \mathcal{F}(\omega,Q) = 2 \pi T^2 \left(e^{-T^2 (\omega - Q)^2}\rho_{11,d} + e^{-T^2 (\omega + Q)^2}\rho_{00,d} \right).
\end{align}

\begin{figure}[t]
    \centering
    \includegraphics[width=0.95\linewidth]{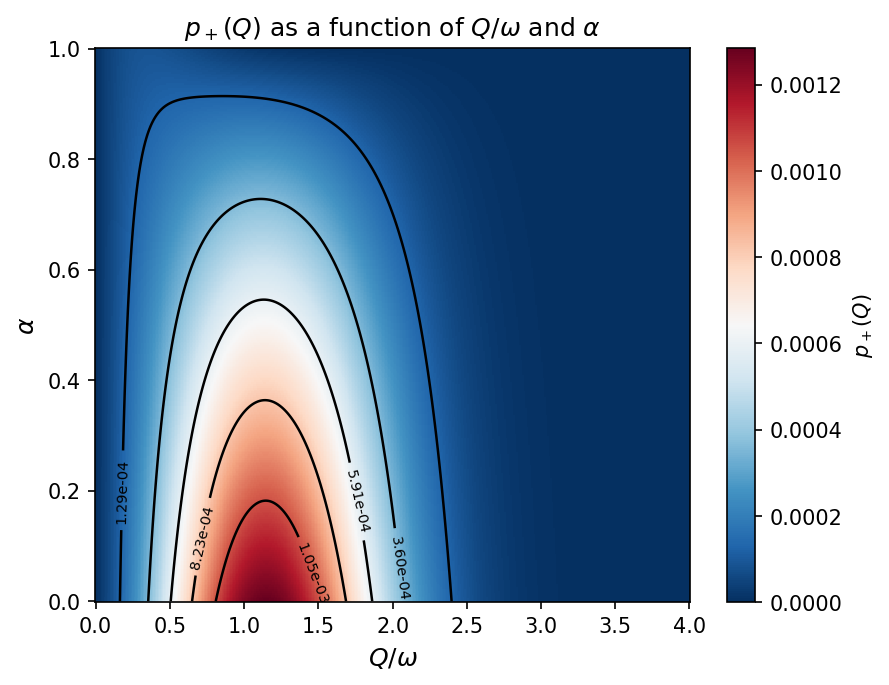}
    \includegraphics[width=0.95\linewidth]{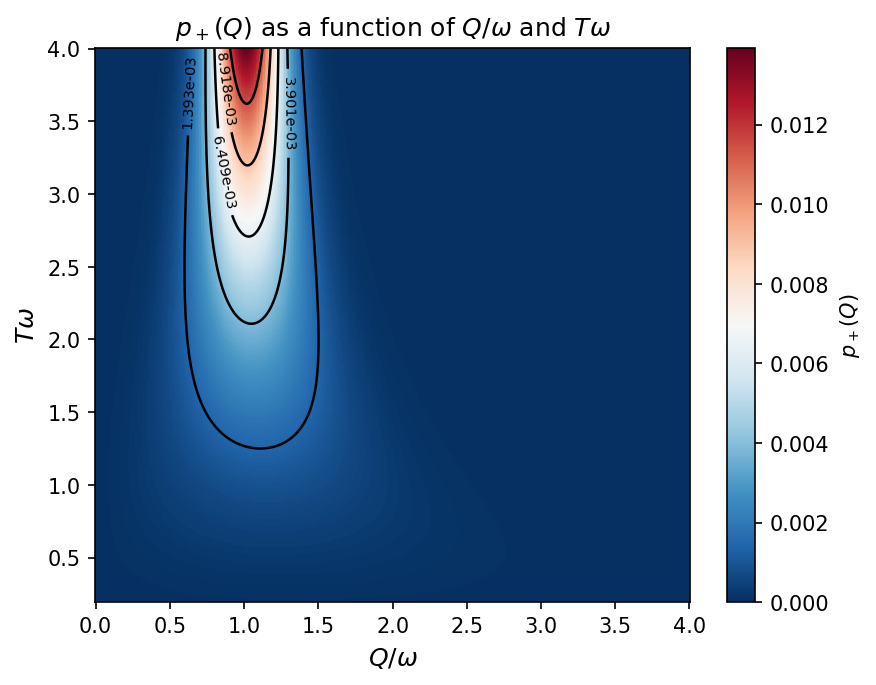}
    \caption{Top panel: Dimensionless probability distribution $p_+(Q>0)=\omega P_+(Q>0)$ as a function of the rescaled heat $Q/\omega$ and the detector's initial population parameter $\alpha=\rho_{00,d}$, for $\lambda=0.1$, $T\omega=1$, and $\sigma\omega=0.5$. Bottom panel: Dimensionless probability distribution $p_+(Q>0)=\omega P_+(Q>0)$ as a function of the rescaled heat $Q/\omega$ and the interaction timescale $T\omega$, for $\lambda=0.1$, $\alpha=0.3$, and $\sigma\omega=0.5$.}
    \label{Fig:Pqvacuum}
\end{figure}

In the vacuum, the field contains no thermal excitations. Nevertheless, a nonzero probability of heat exchange still appears due to the interaction between the detector and quantum vacuum fluctuations. In particular, when the detector is initially prepared in the excited state ($\alpha \approx 0$), there is a significant probability for the detector to undergo a transition to the ground state while emitting energy into the field. This process corresponds to spontaneous emission and leads to a positive heat transfer $Q>0$. As expected, the heat exchange is centered around the detector energy gap, as can be seen from the panels of Fig.~\ref{Fig:Pqvacuum}. As $\alpha$ increases and the detector becomes more likely to start in the ground state, the probability of such transitions decreases. In the limiting case where the detector is initially in its ground state (\(\alpha=1\)), the resonant emission contribution vanishes. Nevertheless, a residual positive-heat contribution remains at finite interaction time due to the counter-rotating term $|\tilde{\chi}(\omega+\omega_k)|^2$. This contribution is suppressed as the interaction time increases and vanishes in the long-time limit. Furthermore, the probability of heat exchange increases with the duration of the interaction. As the interaction time-scale becomes longer, the detector has more time to undergo transitions, leading to a sharper and more pronounced distribution around $Q \sim \omega$. 

This also can be seen by computing the average heat given by Eq.~\eqref{eq:aveQvac} in the long-time interaction limit. In this regime, the switching function effectively enforces energy conservation through
\begin{align}
    \lim_{T \to \infty} |\tilde{\chi}(\omega - k)|^2 = 2 \pi^{3/2} T \, \delta(\omega - k), \label{eq:longtime}
\end{align}
which yields
\begin{align}
  \lim_{T \to \infty} \frac{\langle Q \rangle}{T} = \frac{\lambda^2}{2 \pi^{1/2}} \omega^{2} e^{-\sigma^{2} \omega^{2}} \rho_{11,d}.
\end{align}
This expression shows that, in the long-time limit, the average dissipated heat grows linearly with time, defining a constant emission rate. The delta function enforces the resonant condition $k=\omega$, indicating that the heat exchange is dominated by field modes matching the detector energy gap. The exponential factor $e^{-\sigma^{2} \omega^{2}}$ encodes the suppression due to the spatial smearing of the detector. Moreover, the explicit dependence on $\rho_{11,d}$ shows that only the initially excited population contributes to the emission process. In particular, if the detector starts in the ground state, the average dissipated heat vanishes. Overall, the results show that heat exchange in the vacuum is dominated by decay processes of initially excited detectors, while a detector prepared close to its ground state exhibits a strongly suppressed probability of exchanging energy with the field.

\subsection{KMS-state of inverse temperature $\beta$}
\label{sec:IIIB}
We now turn to the case in which the quantum field is prepared in a thermal KMS state at finite temperature $1/\beta$. The steps required to obtain the characteristic function $C(\nu)$, given by Eq.~\eqref{Eq:Cgeneral}, follow closely those presented in the vacuum case. In particular, the perturbative expansion and operator manipulations remain unchanged, the only modification being the replacement of the vacuum Wightman function $\mathcal{W}_{\text{vac}}$ by its thermal counterpart $\mathcal{W}_{\beta}$. This substitution encodes the presence of thermal excitations in the field and leads to qualitatively new contributions to the heat exchange.

In fact, by using Eq.~\eqref{eq:KMSw} together with Eq.~\eqref{Eq:Cgeneral}, we have
\begin{widetext}
\begin{align}
    C(\nu)  = 1 + \lambda^2 \int \frac{d^3 \textbf{k}}{(2\pi)^32 \omega_{\textbf{k}}} \Big[ \Gamma^+(\omega,\omega_{\textbf{k}},\textbf{k}) (e^{i \nu \omega_{\textbf{k}}} - 1) + \Gamma^-(\omega,\omega_{\textbf{k}},\textbf{k})(e^{-i \nu \omega_{\textbf{k}}} - 1) \Big], \label{eq:CnuKMS}
\end{align}
where
\begin{align}
& \Gamma^+(\omega,\omega_{\textbf{k}},\textbf{k}) = \frac{e^{\beta \omega_{\textbf{k}}}}{e^{\beta \omega_{\textbf{k}}}-1}  \left( |\tilde{\Lambda}(\omega - \omega_{\textbf{k}}, \textbf{k})|^2  \rho_{11,d} + |\tilde{\Lambda}(\omega + \omega_{\textbf{k}}, \textbf{k})|^2 \rho_{00,d}\right), \\
& \Gamma^-(\omega,\omega_{\textbf{k}},\textbf{k}) =  \frac{1}{e^{\beta \omega_{\textbf{k}}}-1}  \left( |\tilde{\Lambda}(\omega - \omega_{\textbf{k}}, \textbf{k})|^2  \rho_{00,d} + |\tilde{\Lambda}(\omega + \omega_{\textbf{k}}, \textbf{k})|^2 \rho_{11,d}\right),
\end{align}
\end{widetext}
with $\tilde{\Lambda}(\omega \pm \omega_{\textbf{k}}, \textbf{k}) = \tilde{\chi}(\omega \pm \omega_{\textbf{k}}) \tilde{\psi}(\textbf{k})$ being the Fourier transform of the spacetime smearing function. Let us notice that, up to $\mathcal{O}(\lambda^2)$, the characteristic function $C(\nu)$ coincides with the low-intensity expansion of a bidirectional Poisson characteristic function~\cite{Hida2008, Esposito2009}. This becomes more transparent by considering the cumulant-generating function $K(\nu) = \ln C(\nu)$ and expanding it to the same order. In this picture, the dynamics can be interpreted as arising from independent stochastic events associated with positive and negative energy exchanges $\pm \omega_{\textbf{k}}$.

By taking the inverse Fourier transform of $C(\nu)$, we have
\begin{align}
P(Q) = & P(Q=0)\delta(Q) + P(Q>0) \delta(Q-\omega_{\mathbf{k}}) \\ & + P(Q<0) \delta(Q+\omega_{\mathbf{k}}) \nonumber,
\end{align}
where
\begin{align}
& P(Q=0) = 1 - P(Q>0) - P(Q<0), \\
& P(Q>0) = \lambda^2
\int \frac{d^3 \mathbf{k}}{(2\pi)^3(2\omega_{\mathbf{k}})}\Gamma^+(\omega,\omega_{\textbf{k}},\textbf{k}), \label{eq:PQpos} \\
& P(Q<0) = \lambda^2
\int \frac{d^3 \mathbf{k}}{(2\pi)^3(2\omega_{\mathbf{k}})}\Gamma^-(\omega,\omega_{\textbf{k}},\textbf{k}) \label{eq:PQneg}, 
%& \mathcal{F}(\mathbf{k}) = |\tilde{\chi}(\omega - \omega_{\textbf{k}})|^2\rho_{11,D} + |\tilde{\chi}(\omega + \omega_{\textbf{k}})|^2\rho_{00,D}, \\ & \mathcal{G}(\mathbf{k}) = |\tilde{\chi}(\omega - \omega_{\textbf{k}})|^2\rho_{00,D} + |\tilde{\chi}(\omega + \omega_{\textbf{k}})|^2\rho_{11,D},
\end{align}
Let us notice that, in contrast to the vacuum state, the heat distribution now contains three contributions: (i) no heat exchange, with probability $P(Q=0)$; (ii) heat absorption by the field, with probability $P(Q>0)$; and (iii) heat release by the field, with probability $P(Q<0)$. Moreover, in the limit $\beta \to \infty$, we recover the results for the vacuum in the previous section and one can easily see that $\int_{-\infty}^{\infty} dQ P(Q) = 1$. 

For $\nu = i\beta$ in Eq.~\eqref{eq:CnuKMS}, the fluctuation relation~\eqref{eq:flucrel} becomes
\begin{widetext}
\begin{align}
\langle e^{-\beta Q} \rangle = C(i\beta) = 1 + \lambda^2 \int \frac{d^3 \textbf{k}}{(2\pi)^3} \frac{|{\tilde{\psi}(\textbf{k})}|^2}{2 \omega_{\textbf{k}}}\left(|\tilde{\chi}(\omega - \omega_{\textbf{k}})|^2 - |\tilde{\chi}(\omega + \omega_{\textbf{k}})|^2\right)(\rho_{00,d} - \rho_{11,d}). \label{eq:LandauerKMS} 
\end{align}    
\end{widetext}
One can see that this quantity is, to leading perturbative order, essentially independent of the temperature of the field, remaining very close to unity within the regime considered here. This behavior is consistent with the emergence of an effective fluctuation relation, in the sense discussed in Ref.~\cite{Taranto}. In that framework, the near-unity value arises from concentration of measure, a phenomenon in high-dimensional systems whereby relevant observables become sharply peaked around their typical values, with fluctuations strongly suppressed as the dimension increases. In particular, for large environments, quantities such as the characteristic function become effectively independent of microscopic details, including the specific initial state, and approach universal values. In our case, the field possesses infinitely many degrees of freedom, which is qualitatively consistent with the large-dimension limit underlying those results. However, the near-unity value already emerges at leading order in perturbation theory and does not rely explicitly on typicality arguments. Instead, it can be traced back to the perturbative structure of the dynamics and the weak, localized coupling between the detector and the field. Furthermore, we observe a residual dependence on the initial state of the detector. As a consequence, although the field has infinitely many degrees of freedom, the conditions required for full concentration of measure are not completely realized, and deviations from universal typical behavior persist. These features will be further assessed numerically when restricting to specific examples. Finally, in the gapless-detector limit, this relation becomes exact, as we will show. In this regime, the absence of an intrinsic detector energy scale removes the spectral selectivity associated with a finite gap, allowing the characteristic function to inherit the symmetry properties encoded in the KMS condition directly.

On the other hand, the average heat exchange can be calculated exactly from the first moment of the characteristic function, i.e., from Eq.~\eqref{Eq:aveQ} we have
\begin{align}
 \langle Q \rangle = \frac{\lambda^2}{2} \int \frac{d^3 \mathbf{k}}{(2\pi)^3} \left(\Gamma^+(\omega,\omega_{\textbf{k}},\textbf{k})  - \Gamma^-(\omega,\omega_{\textbf{k}},\textbf{k}) \right)
\end{align}
In this case, the average heat can also take negative values. A negative average heat, $\langle Q \rangle <0$, indicates that the system extracts heat from the environment, corresponding to a cooling process. %Moreover, from Eq.~\eqref{eq:LandauerKMS}, we see that $- \ln C(i\beta)$ can also become negative. However, the average heat $\langle Q \rangle$ must still satisfy the inequality~\eqref{eq:Landauer} and therefore remains bounded from below by this quantity. 

Finally, it is worth notice that, when the detector is initially prepared in a maximally mixed state, $\rho_{00,d} = \rho_{11,d} = 1/2$, the characteristic function~\eqref{eq:CnuKMS} obeys the fluctuation symmetries $C(-\nu + i\beta) = C(\nu)$ and $C(i \beta) = 1$ exactly,  which implies the detailed fluctuation relation for the exchanged heat of the form $P(Q)/P(-Q) = e^{\beta Q}$ and the Jarzynski equality $\langle e^{-\beta Q}\rangle = 1$. This can be understood from the fact that, in this case, the contributions from excitation and de-excitation processes are equally weighted, eliminating any asymmetry associated with the detector's initial preparation. As a result, the characteristic function inherits directly the symmetry properties of the thermal Wightman function, and the fluctuation relations emerge solely from the KMS condition.

\subsubsection{Massless scalar field}
To gain further insight, we first restrict our analysis to the case of a massless scalar field, for which $\omega_{\mathbf{k}} = ||\mathbf{k}|| = k$. At this stage, we keep the discussion general by considering a class of spatial smearing functions satisfying $\tilde{\psi}(\mathbf{k}) = \tilde{\psi}(k)$, i.e., depending only on the magnitude of the momentum. Specific choices, such as Gaussian spacetime smearing, will be introduced later when performing the numerical analysis.

Let us focus on the positive heat contribution and consider the probability $P_+(Q) = P(Q>0)\,\delta(Q-k)$. From Eq.~\eqref{eq:PQpos} for a fixed amount of exchanged heat, $Q = q > 0$, we obtain
\begin{align}
P_+(q) = \frac{\lambda^2q}{4 \pi^2} \Gamma^+(\omega,q), 
\end{align}
with
\begin{align}
\Gamma^+(\omega,q) = \frac{e^{\beta q}}{e^{\beta q}-1}  \Big( |\tilde{\Lambda}(\omega - q, q)|^2  \rho_{11,d} & \\ + |\tilde{\Lambda}(\omega + q, q)|^2 \rho_{00,d}\Big). \nonumber
\end{align}

Moreover, let us consider the contribution associated with negative heat exchange, described by $P_-(Q) = P(Q<0)\,\delta(Q+k)$. From Eq.~\eqref{eq:PQneg}, for a fixed value of the exchanged heat, $Q = -q < 0$, this expression becomes
\begin{align}
  P_-(-q) = \frac{\lambda^2 q}{4 \pi^2} \Gamma^-(\omega,q),
\end{align}
with
\begin{align}
\Gamma^-(\omega,q) = \frac{1}{e^{\beta q}-1}  \Big( |\tilde{\Lambda}(\omega - q, q)|^2  \rho_{00,d} & \\ + |\tilde{\Lambda}(\omega + q, q)|^2 \rho_{11,d}\Big). \nonumber
\end{align}
implying that
\begin{align}   
\frac{P_+(q) }{P_-(-q)} = \frac{\mathcal{F}(q)}{\mathcal{G}(q)}e^{\beta q},
\end{align}
where
\begin{align}
    & \mathcal{F}(q) = |\tilde{\chi}(\omega - q)|^2\rho_{11,d} + |\tilde{\chi}(\omega + q)|^2\rho_{00,d}, \\ & \mathcal{G}(q) = |\tilde{\chi}(\omega - q)|^2\rho_{00,d} + |\tilde{\chi}(\omega + q)|^2\rho_{11,d}.
\end{align}
This relation has a similar structure to a detailed fluctuation relation for heat exchange between the detector and the field. However, in the present case, this structure is modified by the additional factor $\mathcal{F}(q)/\mathcal{G}(q)$, which encodes features of the detector–field interaction, such as the switching function. Hence, while the exponential term reflects the underlying thermal nature of the field, deviations from the conventional detailed balance condition arise from the localized and finite-time nature of the coupling.

Moreover, one can go a step further by analyzing the long-time interaction limit, in which, from Eq.~\eqref{eq:longtime} with $\omega = k = q$, implies that
\begin{align}
   \lim_{T \to \infty} \frac{P_+(q) }{P_-(-q)} = \frac{\rho_{11,d}}{\rho_{00,d}} e^{\beta q}.  
\end{align}
This result shows that the deviations from the standard detailed fluctuation relation are entirely captured by the detector's initial preparation, while the effects of switching become subleading. Now,
if the detector is initially prepared in a thermal equilibrium state with inverse temperature $\beta_D \neq \beta$, such that $\rho_{11,d} = e^{-\beta_D \omega}\rho_{00,d}$, one finds
\begin{align}
\lim_{T \to \infty} \frac{P_+(q)}{P_-(-q)} =  e^{(\beta - \beta_D)q}. \label{eq:flucrelheat}
\end{align}
which takes the standard form of a detailed fluctuation relation for heat exchange. This result closely parallels the fluctuation relation derived by Ref.~\cite{Jarzynski2004}, where the authors consider two finite-dimensional systems, each initially prepared in a Gibbs state at different temperatures and brought into thermal contact. In that setting, the heat exchange statistics satisfy a detailed fluctuation theorem governed by the temperature difference between the two systems. In contrast, the present scenario involves a quantum field in thermal equilibrium described by a KMS state, interacting with a detector that is itself in a Gibbs state at a different temperature. However, it is important to emphasize that $P_-(-q)$ does not correspond to the time-reversed dynamics of a closed system, as in the original framework considered by Ref.~\cite{Jarzynski2004}. Instead, it should be interpreted as the probability associated with the reverse process, namely the absorption of heat by the detector. In this sense, the relation obtained here is closer in spirit to fluctuation relations formulated for open quantum systems, where forward and backward processes are defined operationally in terms of energy exchange events rather than strict microscopic time-reversal~\cite{Ramezani}. Moreover, in the particular case $\beta = \beta_D$, corresponding to global thermal equilibrium, the ratio~\eqref{eq:flucrelheat} reduces to unity, indicating that forward and backward processes occur with equal probability.

\subsubsection{Numerical analysis of heat distribution}
Let us now specialize to specific choices for the switching and spatial smearing functions, given by Eqs.~\eqref{eq:gauss_smearing} and~\eqref{eq:gauss_switching}, in order to perform the numerical analysis. From this choices, we have
\begin{align}
P_+(q) = \frac{\lambda^2 T^2}{2 \pi} \frac{q e^{-\sigma^2 q^2}}{1 - e^{-\beta q}} \Big( & e^{-T^2 (\omega - q)^2}\rho_{11,d}   \\ & + e^{-T^2 (\omega + q)^2}\rho_{00,d} \Big), \nonumber \\
P_-(-q) =  \frac{\lambda^2 T^2}{2 \pi} \frac{q e^{-\sigma^2 q^2}}{e^{\beta q} - 1}\Big( & e^{-T^2 (\omega - q)^2}\rho_{00,d}  \\ & + e^{-T^2 (\omega + q)^2}\rho_{11,d} \Big). \nonumber
\end{align}

\begin{figure}[t]
    \centering 
    \includegraphics[width=0.75\linewidth]{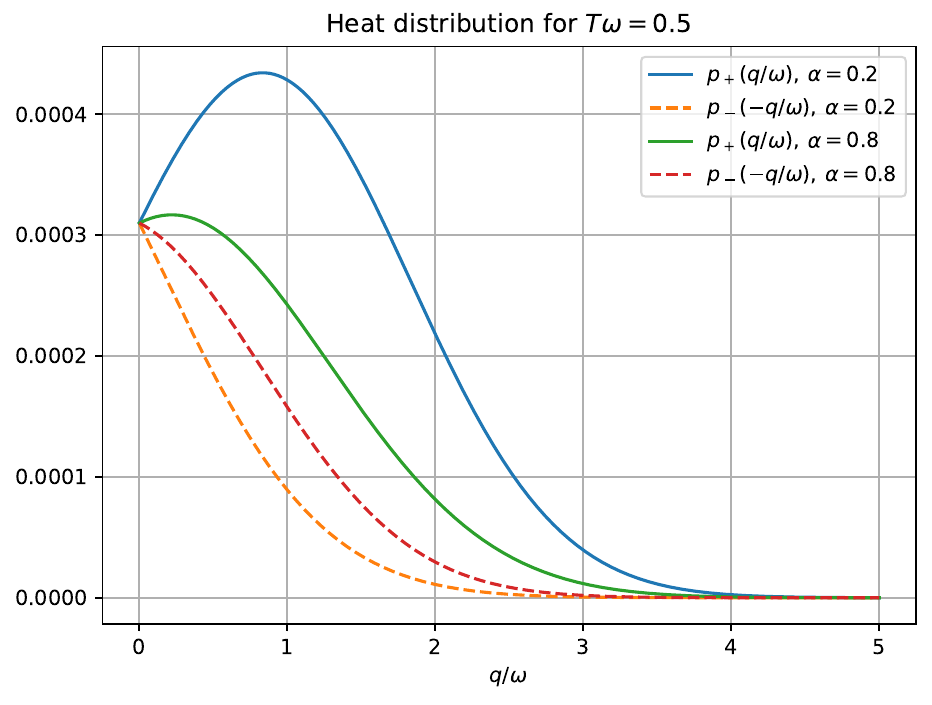}
    \includegraphics[width=0.75\linewidth]{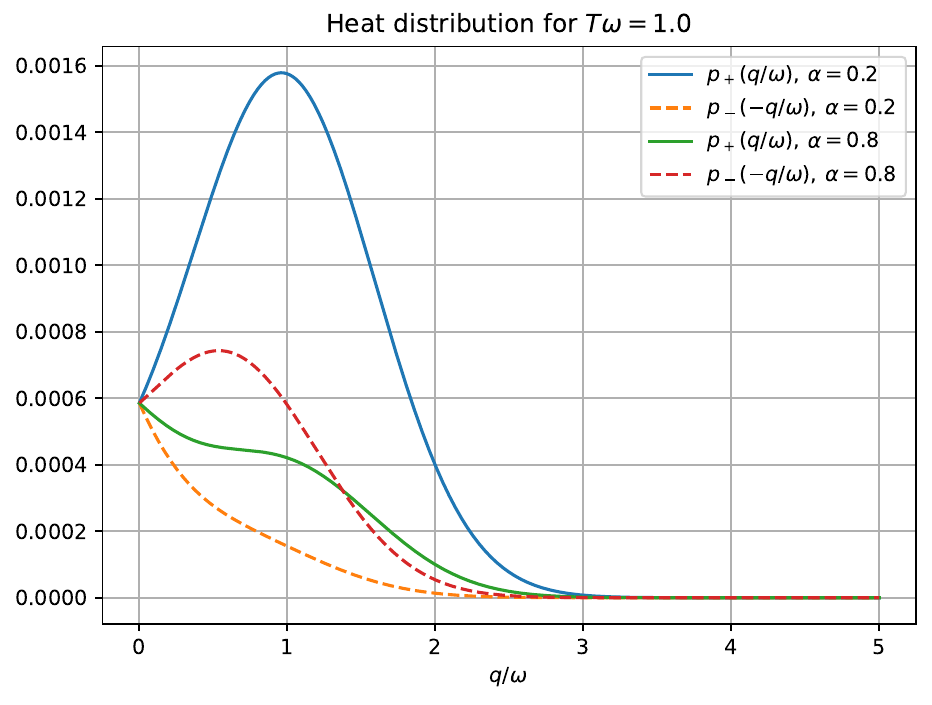}
    \includegraphics[width=0.75\linewidth]{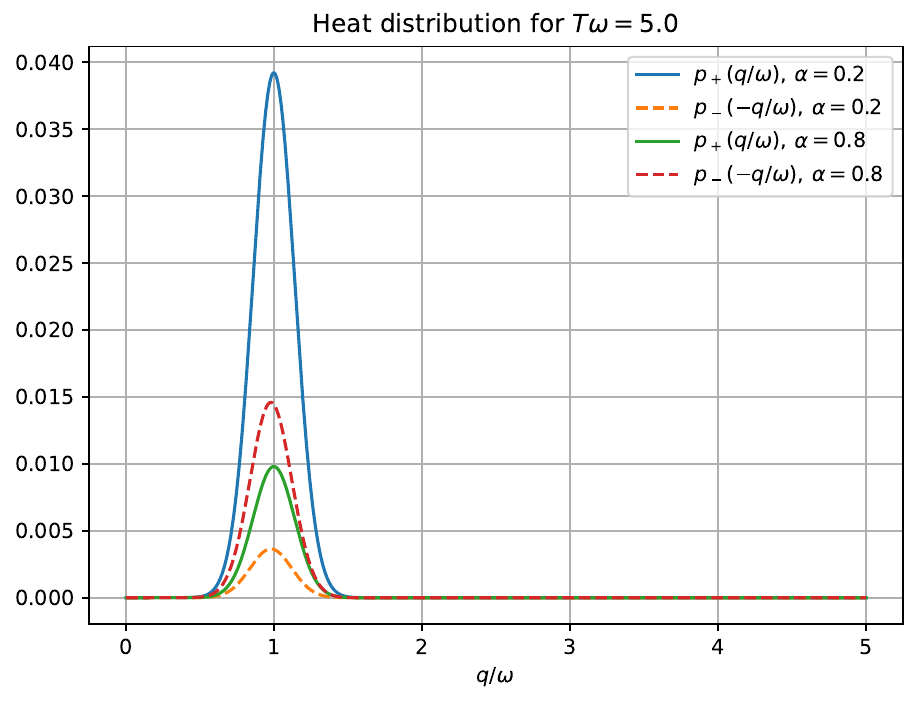}
    \caption{Probability distributions $p_+ = \omega P_+(q/\omega)$ and $p_- = \omega P_-(-q/\omega)$ corresponding respectively to positive and negative heat exchanges between the detector and the field, for $\lambda = 0.1$, $\sigma\omega = 0.5$, $\beta \omega = 1$ and different initial states of the detector, parametrized by $\alpha = \rho_{00,d}$. The panels correspond to different interaction times: top panel $T\omega=0.5$, middle panel $T\omega=1$, and bottom panel $T\omega=5$.}
    \label{fig:Pqdistribution}
\end{figure}

In Fig.~\ref{fig:Pqdistribution}, we display the heat probability distributions $p_+ = \omega P_+(q/\omega)$ and $p_- = \omega P_-(-q/\omega)$, associated with positive and negative energy exchange between the detector and the field. The results are shown for $\lambda = 0.1$, $\sigma \omega = 0.5$, and $\beta \omega = 1$, and for different initial states of the detector, parametrized by $\alpha = \rho_{00,d}$. The three panels correspond to increasing interaction times, namely $T \omega = 0.1$, $T \omega = 1$,  and $T \omega = 5$. The behavior displayed in Fig.~\ref{fig:Pqdistribution} reflects the interplay between the detector's initial state and the interaction timescale. For short interaction times $T \omega=0.1$, both $p_+$ and $p_-$ are broadly distributed, indicating that energy exchange is weak and not sharply resolved in frequency. As the interaction time increases, the distributions become progressively more peaked around $q \sim \omega$, signaling that the detector preferentially exchanges energy quanta close to its gap. This is consistent with the emergence of an approximate energy-conserving regime in the long-time limit. Moreover, the relative weight between $p_+$ and $p_-$ depends on the initial state of the detector. For smaller values of $\alpha$ (detector initially more excited), the positive heat contribution $p_+$ dominates, corresponding to emission into the field. Conversely, for larger $\alpha$ (detector closer to the ground state), the negative heat contribution $p_-$ becomes more significant, reflecting absorption processes from the thermal field.

In addition, one can observe a qualitative change across interaction timescales. For short interaction times, the distributions satisfy $p_+ \ge p_-$ over the range of $q/\omega$, independently of the detector's initial state. However, for intermediate interaction times $T\omega=1$, there exist regions in $q/\omega$ where $p_- > p_+$, particularly when $\alpha \gtrsim 0.5$, i.e., when the detector is more likely to absorb energy from the field. This intermediate regime highlights the nontrivial competition between emission and absorption processes before the long-time limit is reached. In the long-time regime, as the interaction time increases, $p_+$ dominates over $p_-$ when $\alpha < 1/(1 + e^{-\beta \omega})$, while $p_-$ dominates over $p_+$ when $\alpha > 1/(1 + e^{-\beta \omega})$. This threshold corresponds to the thermal population of the detector at inverse temperature $\beta$, signaling the onset of detailed balance in the long-time limit. Overall, Fig.~\ref{fig:Pqdistribution} illustrates how the combined effects of interaction time, detector preparation, and thermal fluctuations shape the statistics of heat exchange.

\begin{figure}
    \centering
    \includegraphics[width=1\linewidth]{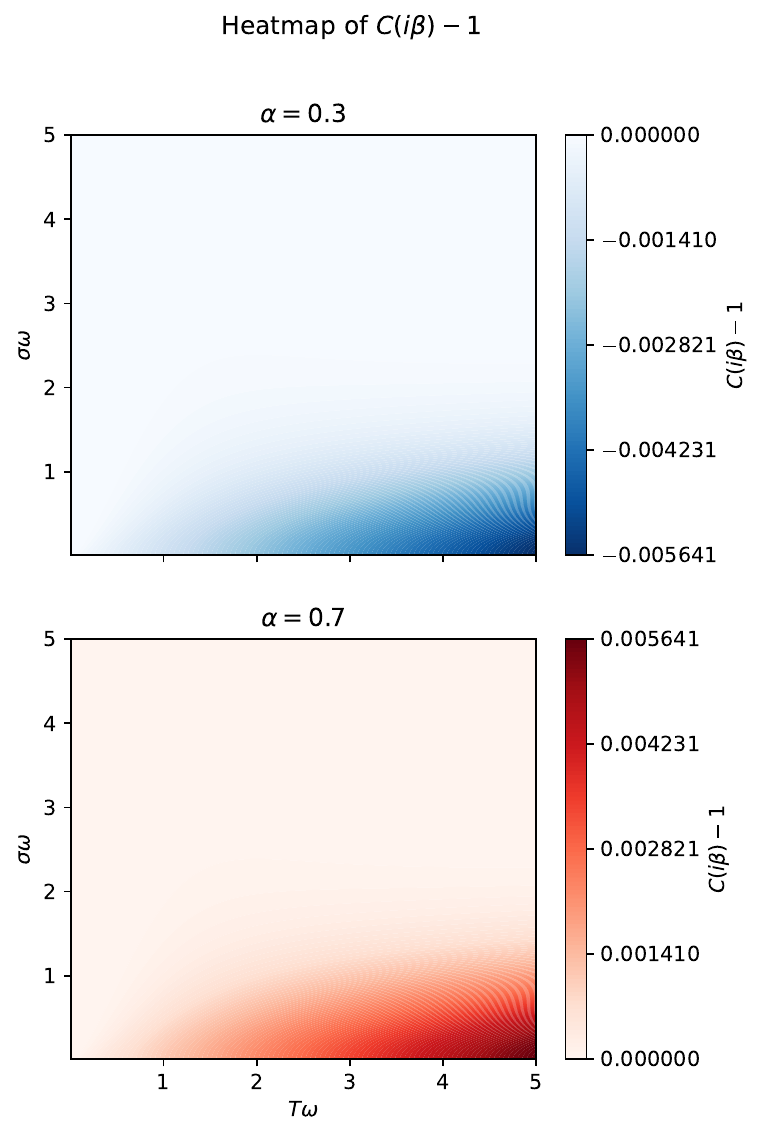}
    \caption{Heatmaps of $C(i\beta)-1$ as a function of the interaction time $T\omega$ and spatial smearing scale $\sigma\omega$, for $\lambda = 0.1$. The top and bottom panels correspond to different initial states of the detector, parametrized by $\alpha = \rho_{00,d}$, with $\alpha = 0.3$ and $\alpha = 0.7$, respectively. The color scale is centered around zero to highlight deviations from the fluctuation relation condition $C(i\beta)=1$.}
    \label{fig:Cibeta}
\end{figure}

Now, the fluctuation relation~\eqref{eq:flucrel} reduces to
\begin{align}
\langle e^{-\beta Q} \rangle   = 1 +  \frac{\lambda^2 T^2}{2 \pi} & \Big[ \frac{T^2 \omega \sqrt{\pi}}{(\sigma^2 + T^2)^{3/2}} \\ & \times e^{-(\frac{\sigma^2 T^2}{\sigma^2 + T^2})\omega^2}\Big](\rho_{00,d} - \rho_{11,d}). \nonumber
\end{align}
From the expression above, we present heatmaps of $C(i\beta)-1$ in Fig.~\ref{fig:Cibeta} as a function of the interaction time $T$ and the spatial smearing scale $\sigma$, for different initial states of the detector. As anticipated from the perturbative analysis, $\langle e^{-\beta Q}\rangle =C(i\beta)$ remains close to unity over a wide range of parameters, reinforcing the emergence of an effective fluctuation relation in this regime.
Deviations from unity, although small, exhibit a clear dependence on the detector's initial state. In particular, we observe that $\langle e^{-\beta Q}\rangle =C(i\beta)$ can become smaller than one, which is typically associated with regimes where the detector predominantly releases energy into the field, i.e., when emission processes dominate the dynamics. Conversely, values slightly above unity correspond to situations where absorption from the field becomes more relevant.

Moreover, from Eq.~\eqref{Eq:aveQ}, the averaged heat $\langle Q \rangle$ can be computed from the integral
\begin{align}
     \langle Q \rangle =
\frac{\lambda^{2} T^{2}}{2\pi}
\int_{0}^{\infty} dk\, & k^{2} e^{-\sigma^{2} k^{2}}
\Big[ 
\frac{\rho_{11,d} e^{\beta k} - \rho_{00,d}}{e^{\beta k}-1}
e^{-T^{2}(\omega-k)^{2}}
\\ & +
\frac{\rho_{00,d} e^{\beta k} - \rho_{11,d}}{e^{\beta k}-1}
e^{-T^{2}(\omega+k)^{2}}
\Big]. \nonumber
\end{align}

Before proceeding with the numerical analysis, it is instructive to examine the long-time interaction limit. From Eq.~\eqref{eq:longtime}, one obtains
\begin{align}
  \lim_{T \to \infty} \frac{\langle Q \rangle}{T} = \frac{\lambda^2}{2 \pi^{1/2}} 
\frac{\omega^{2} e^{-\sigma^{2} \omega^{2}}}{e^{\beta \omega}-1}(\rho_{11,d} e^{\beta \omega} - \rho_{00,d}).
\end{align}
This expression shows that, in the long-time limit, the average heat scales linearly with the interaction time, and the corresponding rate is determined by the competition between emission and absorption processes encoded in the combination $\rho_{11,d} e^{\beta \omega} - \rho_{00,d}$. In particular, the sign of the heat flow depends on the initial population of the detector. When $\rho_{11,d} e^{\beta \omega} > \rho_{00,d}$, emission processes dominate and the detector transfers energy to the field, leading to positive heat dissipation. Conversely, when $\rho_{11,d} e^{\beta \omega} < \rho_{00,d}$, absorption processes prevail and the detector extracts energy from the field. A particularly relevant case occurs when the detector is initially prepared in a thermal state at the same inverse temperature $\beta$ as the field. In this situation, the detailed balance condition $\rho_{11,d}/\rho_{00,d} = e^{-\beta \omega}$ implies
\begin{align}
    \lim_{T \to \infty} \frac{\langle Q \rangle}{T} = 0,
\end{align}
indicating that no net heat is exchanged in the steady regime. This result is consistent with thermodynamic equilibrium, where emission and absorption processes exactly compensate each other, leading to vanishing average heat flow despite the presence of ongoing microscopic transitions.

\begin{figure}[t]
    \centering
    \includegraphics[width=0.99\linewidth]{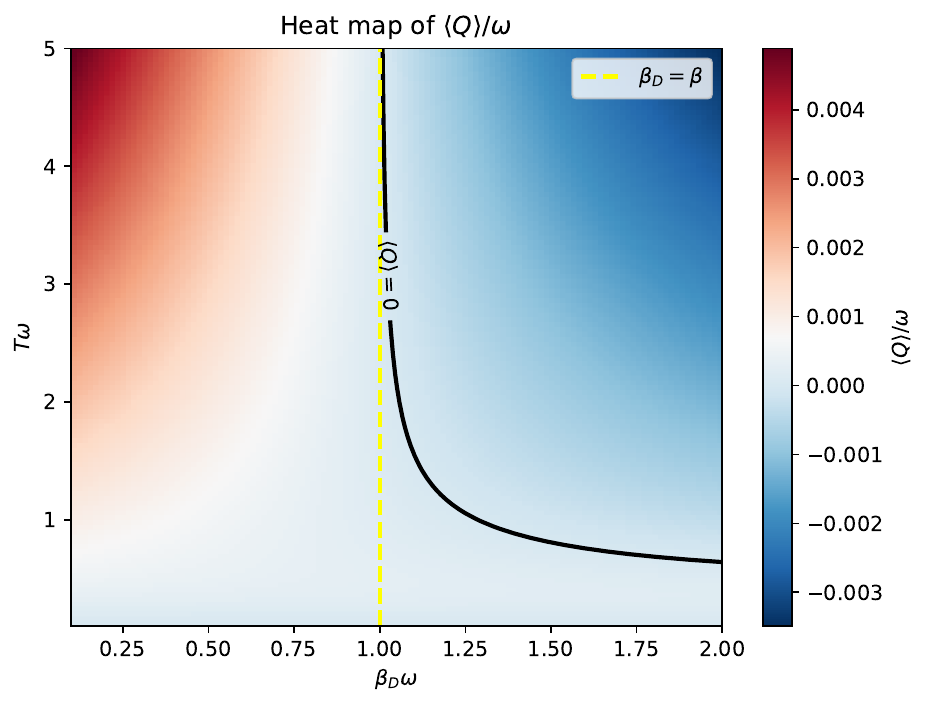}
    \caption{Heatmap of the average heat exchanged $\langle Q \rangle/\omega$ as a function of the interaction time $T\omega$ and the inverse temperature of the detector $\beta_D \omega$, for $\lambda = 0.1$, $\sigma \omega = 0.5$, and a field prepared in a KMS state at inverse temperature $\beta \omega = 1$. The black curve indicates the locus where $\langle Q \rangle/\omega = 0$, separating regimes of net heat dissipation $\langle Q \rangle /\omega > 0$ and absorption $\langle Q \rangle /\omega < 0$. The dashed vertical line marks the equilibrium condition $\beta_D = \beta$.}
    \label{fig:averageheatmap}
\end{figure}

To further investigate the thermodynamic behavior between the detector and the field, in Fig.~\ref{fig:averageheatmap}, we show the average heat exchanged with the thermal field, $\langle Q \rangle/\omega$, as a function of the interaction time $T\omega$ and of the inverse temperature of the detector $\beta_D\omega$, assuming that the detector is initially prepared in a thermal state at temperature $\beta_D^{-1}$. The results are shown for $\lambda = 0.1$, $\sigma \omega = 0.5$, and a field prepared in a KMS state at inverse temperature $\beta \omega = 1$. The black curve indicates the locus where $\langle Q \rangle/\omega = 0$, separating two distinct thermodynamic regimes. The zero-heat contour appears as a single connected curve, indicating a unique crossover between emission and absorption regimes across the explored parameter space. For sufficiently long interaction times, the direction of heat flow follows the expected thermodynamic behavior: for $\beta_D < \beta$, the detector is initially hotter than the field and dissipates energy into it, leading to $\langle Q \rangle/\omega > 0$, while for $\beta_D > \beta$, the detector is colder and absorbs energy from the thermal bath, resulting in $\langle Q \rangle/\omega < 0$. However, for finite interaction times, deviations from this behavior arise. In particular, even when $\beta_D > \beta$, there exist regions where $\langle Q \rangle/\omega > 0$, reflecting transient effects and the fact that the detector only probes a finite portion of the field degrees of freedom. The dashed vertical line $\beta_D = \beta$ corresponds to the equilibrium condition where the detector and the field share the same temperature. As expected, this line closely follows the zero-heat curve as the interaction time increases. In the long-time limit, these curves coincide, indicating that thermal equilibrium implies vanishing net heat exchange. For finite interaction times, however, small deviations are observed, reflecting the localized and time-dependent nature of the interaction. Overall, Fig.~\ref{fig:averageheatmap} illustrates how both the interaction time and the initial temperature imbalance between the detector and the field govern the direction and magnitude of heat flow, smoothly interpolating between transient and near-equilibrium regimes.

\subsubsection{Numerical comparison of the Landauer bounds}
\label{sec:numerical-landauer-comparison}
We now use the perturbative detector dynamics to compare the numerical values of the lower bounds in Eqs.~\eqref{eq:Landauer} and \eqref{eq:Landauer1}. Their validity has already been established analytically. The purpose of the calculation is to determine their relative tightness for the UDW detector–field interaction considered here.

The evolved reduced density-matrix of the particle detector is given by
\begin{align}
    \hat{\rho}_{d,\tau} =  \Tr_{\phi}  \left(\hat{U}_I(\tau)  \hat{\rho}_{d,0} \otimes \hat{\rho}_{\beta}  \hat{U}^{\dagger}_I(\tau)\right).
\end{align}
Up to $\mathcal{O}(\lambda^2)$, we have
\begin{align}
\hat{\rho}_{d,\tau} = \hat{\rho}_{d,0} + \lambda^2\delta \hat{\rho}_{d,\tau}, \label{eq:evolvedrho}
\end{align}
such that
\begin{align}
& \langle 0 | \delta \hat{\rho}_{d,\tau} | 0 \rangle = \mathcal{L}^+\langle 1 |\hat{\rho}_{d,0}  | 1 \rangle - \mathcal{L}^-\langle 0 |\hat{\rho}_{d,0}  | 0 \rangle , \\
& \langle 1 | \delta \hat{\rho}_{d,\tau} | 1 \rangle = \mathcal{L}^-\langle 0 |\hat{\rho}_{d,0}  | 0 \rangle - \mathcal{L}^+ \langle 1 |\hat{\rho}_{d,0}  | 1 \rangle, \\
& \langle 0 | \delta\hat{\rho}_{d,\tau} | 1 \rangle =  \mathcal{K}\langle 1 |\hat{\rho}_{d,0}  | 0 \rangle - \mathcal{N} \langle 0 |\hat{\rho}_{d,0}  | 1 \rangle, 
\end{align}
with 
\begin{align}
& \mathcal{L}^+ =  \iint d^4 \mathsf{x} d^4 \mathsf{x}'\Lambda(\mathsf{x}) \Lambda(\mathsf{x}') \mathcal{W}_{\beta}(\tau, \tau') e^{i \omega(\tau-\tau')}, \\
& \mathcal{L}^- = \iint d^4 \mathsf{x} d^4 \mathsf{x}' \Lambda(\mathsf{x}) \Lambda(\mathsf{x}') \mathcal{W}_{\beta}(\tau, \tau') e^{- i \omega(\tau-\tau')},\\
&  \mathcal{K} = \iint d^4 \mathsf{x} d^4 \mathsf{x}' \Lambda(\mathsf{x}) \Lambda(\mathsf{x}') \mathcal{W}_{\beta}(\tau, \tau') e^{- i \omega(\tau+\tau')}, \\
& \mathcal{N} = \iint d^4 \mathsf{x} d^4 \mathsf{x}' \Lambda(\mathsf{x}) \Lambda(\mathsf{x}') \mathcal{W}_{\beta}(\tau, \tau') e^{- i \omega|\tau-\tau'|},
\end{align}
Similar expressions for $\delta \hat{\rho}_{d,\tau}$ were obtained in Ref.~\cite{shah2025} for a spin interacting with an electromagnetic field.

\begin{figure}[t]
   \centering
    \includegraphics[width=0.8\linewidth]{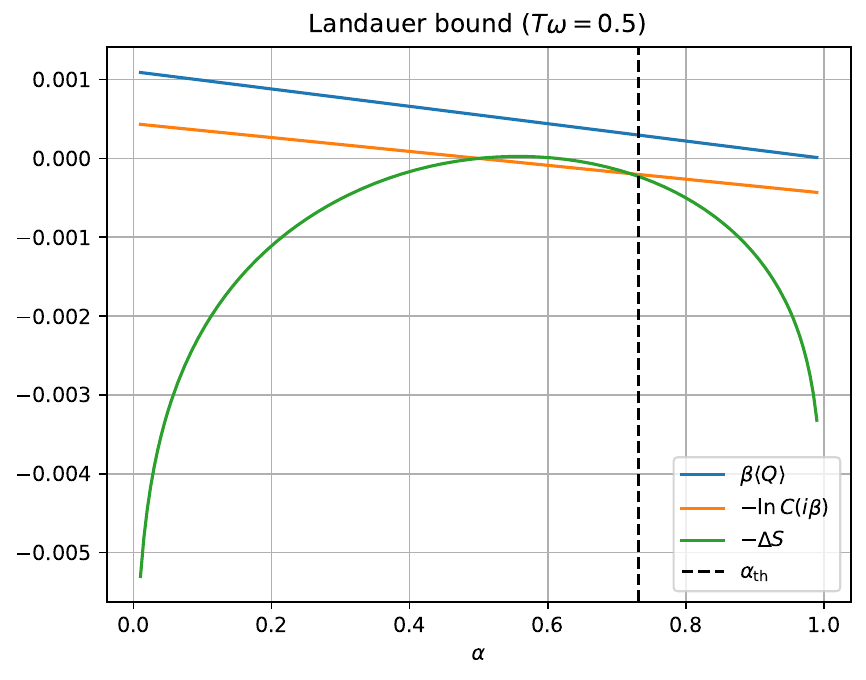}
    \includegraphics[width=0.8\linewidth]{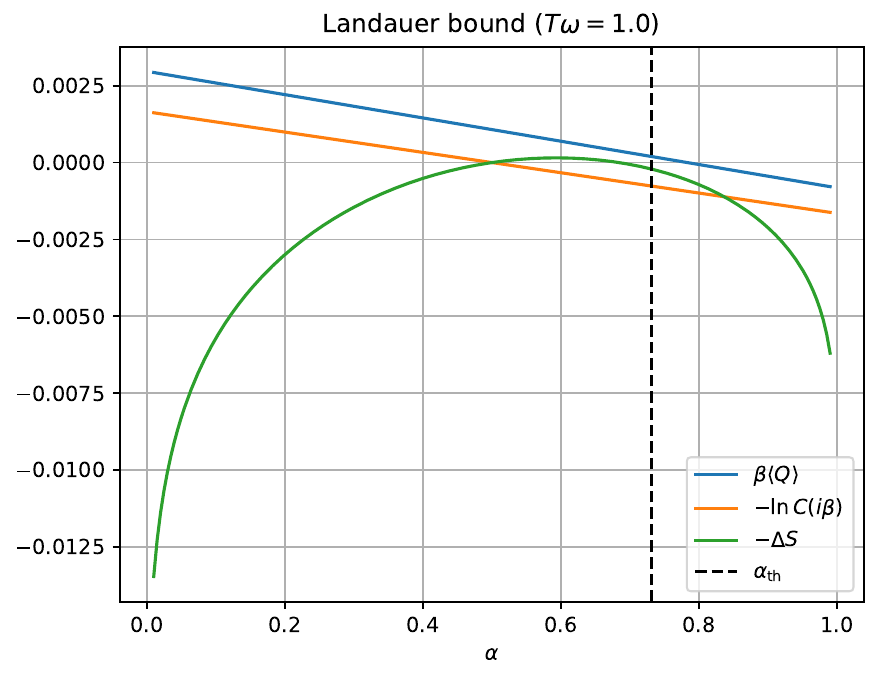}
    \includegraphics[width=0.8\linewidth]{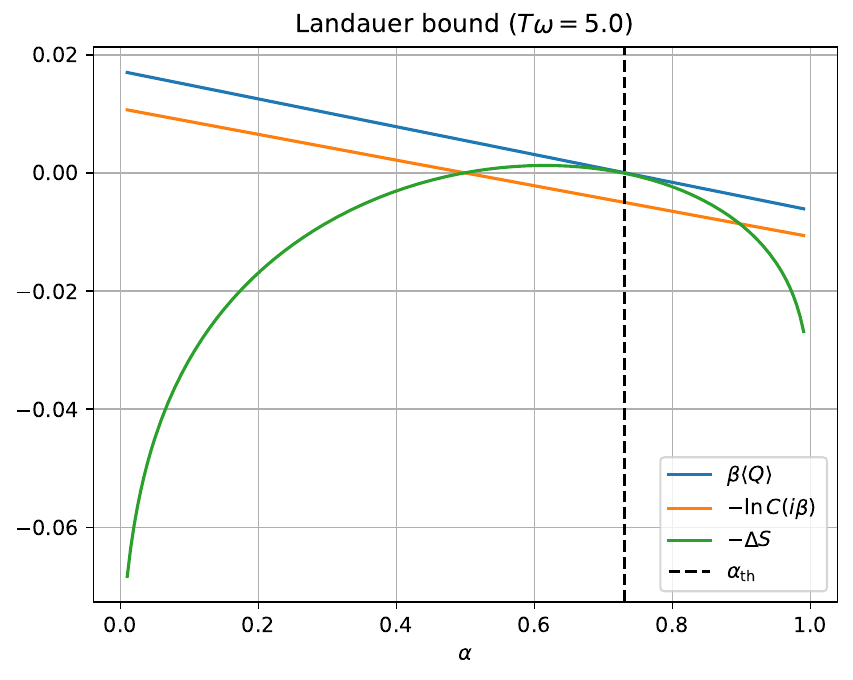}
    \caption{Landauer bound quantities as functions of the initial ground-state population $\alpha = \rho_{00,d}$ for different interaction times $T$. The solid, dashed, and dotted curves correspond to $\beta \langle Q \rangle$, $-\ln C(i\beta)$, and $-\Delta S$, respectively. The vertical dashed line indicates the thermal equilibrium population $\alpha_{\mathrm{th}} = 1/(1+e^{-\beta \omega})$. The results are shown for $\lambda = 0.1$, $\sigma \omega = 0.5$, and $\beta \omega = 1$. Each panel corresponds to a different interaction time $T \omega$, illustrating how the bounds approach each other as the interaction time increases.}
    \label{fig:Landauer}
\end{figure}

We now separate the analysis into two cases: a detector initially prepared in a mixed state with no quantum coherence and a detector initially prepared in a pure state containing coherence. The former leads to simpler calculations, whereas the latter requires a more involved treatment.

\textit{Detector initially prepared in a mixed state with no quantum coherence:} $\langle 1 |\hat{\rho}_{d,0}  | 0 \rangle = 0$ implies that $\langle 0 | \hat{\rho}_{d,\tau} | 1 \rangle = 0$ and the initial state the UDW detector can be taken as $ \hat{\rho}_{d,0} = \alpha |0 \rangle \langle 0| + (1 - \alpha) |1 \rangle \langle 1|$. In this case, we have
\begin{align}
& \langle 0 | \delta \hat{\rho}_{d, \tau} | 0 \rangle  = \mathcal{L}^+(1 - \alpha) - \mathcal{L}^- \alpha , \\
& \langle 1 | \delta \hat{\rho}_{d, \tau} | 1 \rangle =  \mathcal{L}^- \alpha - \mathcal{L}^+(1 - \alpha),
\end{align}
with
\begin{align}
    \mathcal{L}^+ =  \int \frac{d^3 \textbf{k}}{(2\pi)^3 2 \omega_{\textbf{k}}} & \frac{|{\tilde{\psi}(\textbf{k})}|^2}{e^{\beta \omega_{\textbf{k}}}-1}\Big(|\tilde{\chi}(\omega - \omega_{\textbf{k}})|^2 e^{\beta \omega_{\textbf{k}}}\nonumber \\ & + |\tilde{\chi}(\omega + \omega_{\textbf{k}})|^2\Big), \label{eq:L+}\\
 \mathcal{L}^- = \int \frac{d^3 \textbf{k}}{(2\pi)^3 2 \omega_{\textbf{k}}} & \frac{|{\tilde{\psi}(\textbf{k})}|^2}{e^{\beta \omega_{\textbf{k}}}-1}\Big(|\tilde{\chi}(\omega + \omega_{\textbf{k}})|^2 e^{\beta \omega_{\textbf{k}}} \nonumber \\ & + |\tilde{\chi}(\omega - \omega_{\textbf{k}})|^2\Big), \label{eq:L-}
\end{align}
which allows us obtain the probabilities $p_{0,\tau} = \langle 0 | \hat{\rho}_{d,\tau} | 0 \rangle $ and $p_{1,\tau} = \langle 1 | \hat{\rho}_{d,\tau} | 1 \rangle $ and to numerically evaluate the entropy variation $\Delta S = S( \hat{\rho}_{d,\tau}) - S( \hat{\rho}_{d,0})$, where $S(\hat{\rho}_{d,\tau}) = - \Tr \hat{\rho}_{d,\tau} \ln \hat{\rho}_{d,\tau}$, for a massless scalar field with Gaussian spacetime smearing functions given by Eqs.~\eqref{eq:gauss_switching} and~\eqref{eq:gauss_smearing}.

The different Landauer bounds, given by Eq.~\eqref{eq:Landauer} and~\eqref{eq:Landauer1}, are illustrated in Fig.~\ref{fig:Landauer}, where we plot the quantities $\beta \langle Q \rangle$, $-\ln C(i\beta)$, and $-\Delta S$ as functions of the initial ground-state population $\alpha = \rho_{00,d}$ for different interaction times $T\omega$. The results are shown for $\lambda = 0.1$, $\sigma \omega = 0.5$, and $\beta \omega = 1$. We find that the Landauer bounds, given by Eqs.~\eqref{eq:Landauer} and~\eqref{eq:Landauer1}, are satisfied throughout the parameter range considered, and that variations of these fixed parameters do not lead to any violation of the bound within the perturbative regime explored here. As the interaction time increases, the different bounds become progressively closer.

\textit{Detector initially prepared in a pure state containing coherence:} Given $|\psi\rangle_d = \alpha |0\rangle + \sqrt{1-\alpha^2}|1\rangle, \quad \alpha \in [0,1]$, the evolved density matrix of the UDW detector is given by Eq.~\eqref{eq:evolvedrho}
with
\begin{align}
\hat{\rho}_{d,0}=
\begin{pmatrix}
\alpha^2 & \alpha\sqrt{1-\alpha^2} \\
\alpha\sqrt{1-\alpha^2} & 1-\alpha^2
\end{pmatrix},    
\end{align}
and
\begin{align}
    & \langle 0| \delta \hat{\rho}_{d,\tau}|0\rangle = \mathcal{L}^+(1-\alpha^2) - \mathcal{L}^- \alpha^2, \\
    & \langle 1| \delta \hat{\rho}_{d,\tau}|1\rangle = \mathcal{L}^- \alpha^2 - \mathcal{L}^+(1-\alpha^2), \\
    & \langle 0| \delta \hat{\rho}_{d,\tau}|1\rangle = (\mathcal{K}-\mathcal{N})\alpha\sqrt{1-\alpha^2}, \\
    & \langle 0| \delta \hat{\rho}_{d,\tau}|1\rangle = \langle 1| \delta \hat{\rho}_{d,\tau}|0\rangle^*.
\end{align}

From the equations above, the eigenvalues of $\hat{\rho}_{d,\tau}$ can be calculated and are given by
\begin{align}
    \lambda_\pm = \frac{1}{2}\left(1 \pm \sqrt{1 - 4\det \hat{\rho}_{d,\tau}}\right)
\end{align}
with
\begin{align}
    \det\hat{\rho}_{d,\tau}= \lambda^2\Big[ & (2\alpha^2 -1)(\mathcal{L}^- \alpha^2 - \mathcal{L}^+(1-\alpha^2)) \nonumber \\ &  - 2\alpha^2(1-\alpha^2)\mathfrak{R}(\mathcal{K}-\mathcal{N})\Big].
\end{align}
The quantities $\mathcal{L}^{\pm}$ are given by Eqs.~\eqref{eq:L+} and~\eqref{eq:L-}, while
\begin{align}
     \mathfrak{R}(\mathcal{K}-\mathcal{N}) = & \int \frac{d^3 \textbf{k}}{(2\pi)^3 2 \omega_{\textbf{k}}} |{\tilde{\psi}(\textbf{k})}|^2 \coth\frac{\beta \omega_{\textbf{k}}}{2}\nonumber \\& \times
     \Big[\mathfrak{R}\left(\tilde{\chi}^{*}(\omega + \omega_{\textbf{k}})\tilde{\chi}^{*}(\omega - \omega_{\textbf{k}})\right) \\ & - \frac{1}{2}|\tilde{\chi}(\omega + \omega_{\textbf{k}})|^2 - \frac{1}{2}|\tilde{\chi}(\omega - \omega_{\textbf{k}})|^2\nonumber\Big].
\end{align}
which allow us to numerically evaluate the entropy variation $\Delta S = S( \hat{\rho}_{d,\tau}) - S( \hat{\rho}_{d,0})$, where $S(\hat{\rho}_{d,\tau}) = - \Tr \hat{\rho}_{d,\tau} \ln \hat{\rho}_{d,\tau}$, for a massless scalar field with Gaussian spacetime smearing functions given by Eqs.~\eqref{eq:gauss_switching} and~\eqref{eq:gauss_smearing}.

\begin{figure}[t]
   \centering
    \includegraphics[width=0.8\linewidth]{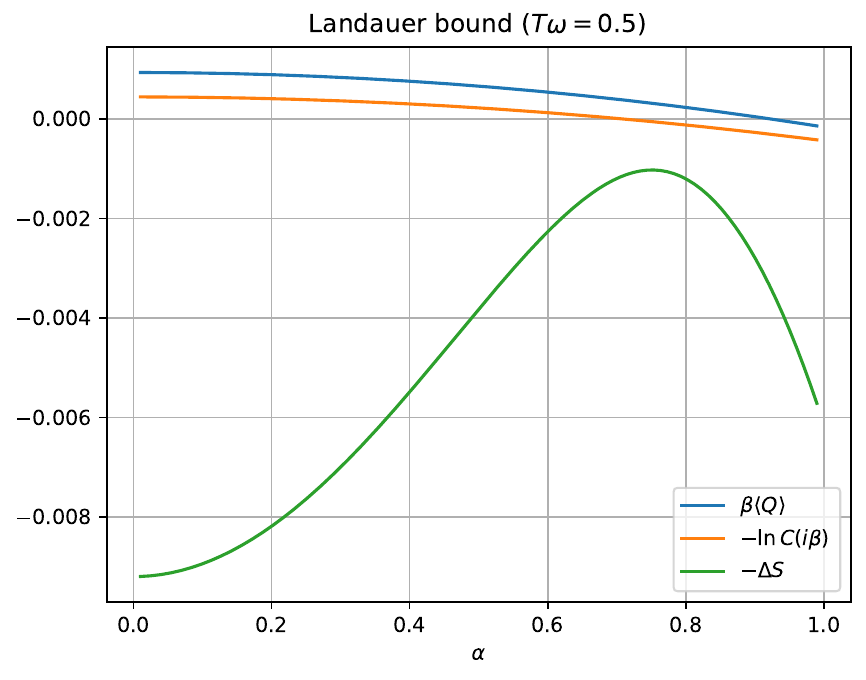}
    \includegraphics[width=0.8\linewidth]{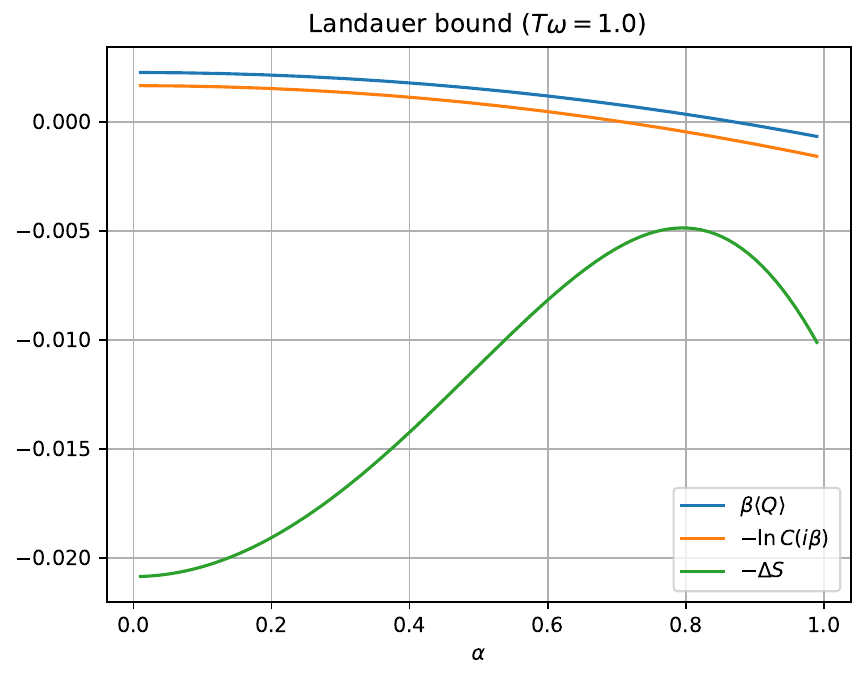}
    \includegraphics[width=0.8\linewidth]{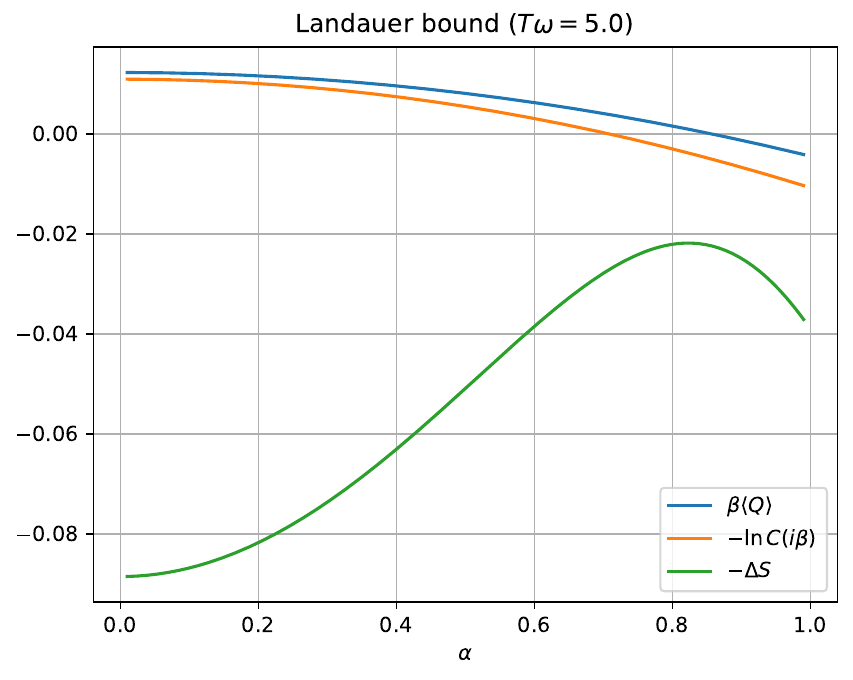}
    \caption{Landauer bound with initial coherence for different interaction times $T$. We plot $\beta\langle Q\rangle$, $-\ln C(i\beta)$, and $-\Delta S$ as functions of the initial state parameter $\alpha$, for $T\omega=0.5$ (top), $T\omega=1$ (middle), and $T\omega=5$ (bottom). The results are shown for $\lambda = 0.1$, $\sigma \omega = 0.5$, and $\beta \omega = 1$.}
    \label{fig:Landauercoherence}
\end{figure}

In Fig.~\ref{fig:Landauercoherence}, we plot the quantities $\beta \langle Q \rangle$, $-\ln C(i\beta)$, and $-\Delta S$ as functions of $\alpha$ for different interaction times $T\omega$. The results are shown for $\lambda = 0.1$, $\sigma \omega = 0.5$, and $\beta \omega = 1$. We find once again that the different forms of the Landauer bound are satisfied for all values of the initial state parameter $\alpha$. One can notice that, in this case, Eq.~\eqref{eq:Landauer1} provides a tighter bound than Eq.~\eqref{eq:Landauer} throughout the entire range of $\alpha$.

We now discuss some general aspects of related works in the literature. First, for the detector with initial zero coherence, our results can be directly compared with those obtained in Ref.~\cite{Hashimoto}, where a two-level system coupled to a bosonic reservoir is analyzed within a full counting statistics framework. In that work, the same two bounds on the average heat are considered. A key result of that analysis is that the relative strength of these bounds depends on the initial state of the system. In particular, the entropic bound~\eqref{eq:Landauer} is tighter for initially mixed states, whereas the bound~\eqref{eq:Landauer1} becomes tighter in the vicinity of pure states. Our results exhibit a closely analogous structure. This behavior is clearly visible in Fig.~\ref{fig:Landauer}, where the crossover between the two bounds depends on the value of $\alpha$. Another important point of agreement concerns the role of coherence. In Ref.~\cite{Hashimoto}, it is shown that the average heat exchanged is independent of the initial coherences of the system. We find the same feature in our model: the mean heat depends only on the initial populations of the detector, while coherence contributes only to off-diagonal terms in the reduced dynamics and does not affect the average energy exchange.

It is worth emphasizing that, although both setups involve bosonic environments, there is an important conceptual distinction between a generic bosonic reservoir and a quantum field. A quantum field possesses additional structure, such as locality, relativistic causality, and spacetime-dependent correlation functions encoded in the Wightman function. In contrast, standard bosonic reservoirs used in open quantum systems are typically characterized only by their spectral density and lack an explicit notion of spacetime localization. In Ref.~\cite{Hashimoto}, the bosonic environment is modeled through an Ohmic spectral density with an exponential cutoff, $J(\omega) = \lambda \omega e^{-\omega/\Omega}$, which effectively encodes all properties of the bath and allows one to control infrared and ultraviolet contributions. In particular, such a choice ensures the positivity of the reduced dynamics within the second-order quantum master equation. In contrast, in the present framework the environment is a quantum field, and the corresponding spectral properties are not adjustable but instead dictated by the underlying field theory. Regularization is introduced through the switching and smearing functions rather than by an explicit cutoff in the spectral density. As a consequence, the dynamical behavior and thermodynamic bounds obtained here are less flexible and more directly constrained by the fundamental structure of the theory. Moreover, while in Ref.~\cite{Hashimoto} the environment is assumed to be in a thermal Gibbs state, here the field is prepared in a KMS state, which provides a more general and intrinsically quantum-field-theoretic notion of thermal equilibrium.

\section{Nonperturbative Evaluation of Heat Statistics}
\label{Sec:IV}
The results obtained so far reveal a rich thermodynamic structure arising from the interaction between the detector and the quantum field, including the validity of Landauer-type bounds, the role of finite-time effects, and deviations from standard equilibrium intuition. However, all these results have been derived within a perturbative framework, valid to leading order in the coupling strength $\lambda$. While this regime already captures several nontrivial features, it does not fully address the behavior of the system beyond weak coupling or allow one to assess the robustness of the observed thermodynamic relations in a non-perturbative setting.

In this context, it is natural to consider a simplified scenario in which analytical progress beyond perturbation theory becomes possible. This is achieved by focusing on a gapless detector ($\omega = 0$) or a delta-switching interaction ($\chi(\tau) \sim \delta(\tau - \tau_0)$). Before proceeding with the calculations, it is instructive to clarify the physical meaning of the gapless detector limit and sharply localized (delta) interaction in the UDW model~\cite{Landulfo2016, Tjoa22, Tjoa2023, Perche2024a, Kasprzak2025}. In the usual setting, particle detectors are employed as operational probes of field quanta, where absorption processes are interpreted as detections of the field quanta. This interpretation relies on the presence of a finite energy gap. Nevertheless, gapless detectors remain valuable probes of quantum fields, as they are particularly sensitive to field correlations. Intuitively, one may think of a detector as a two-level system whose energy gap is induced by an external field. Removing this gap eliminates the intrinsic energy scale of the detector, so that its dynamics is entirely driven by the interaction with the quantum field. As a result, without an intrinsic energy scale, the detector does not selectively couple to specific frequencies, and its response is instead governed directly by the field correlations. In this sense, gapless detectors provide a more direct probe of the structure of quantum fluctuations, rather than of particle content. A closely related scenario is that of a detector interacting with the field through a sharply localized switching. Although conceptually different, this setup leads to an equivalent non-perturbative description. In particular, if one includes the internal dynamics of the detector, their effect reduces to a unitary rotation of the monopole operator $\hat{m}(\tau)$. As a consequence, the presence of an energy gap does not introduce any qualitatively new effects in this regime. In this sense, delta-coupled detectors behave effectively as gapless detectors, independently of their internal Hamiltonian (See Refs.~\cite{Perche2024a, Tjoa2023} for further discussion).

Importantly, both the gapless limit and the delta-switching interaction allow for an exact, non-perturbative treatment of the dynamics. In what follows, we will focus on the gapless detector, since the expression for the characteristic function obtained in both cases is equivalent, a correspondence that will be outlined throughout the discussion. Here, ``non-perturbative'' means that the time-evolution operator can be computed in closed form, capturing contributions to all orders in the coupling strength~\cite{Landulfo2016, Tjoa22}. This feature provides a useful complementary perspective to the perturbative analysis developed previously. Moreover, this regime offers additional physical insight. As the detector becomes gapless, it effectively probes all frequency modes of the field without energy selectivity, which is consistent with the behavior observed in the long-time limit of the finite-gap case. Therefore, the gapless detector can be interpreted as an idealized scenario in which the detector fully samples the field degrees of freedom, providing a complementary perspective on the emergence of fluctuation relations.

We now begin by considering the case of a gapless UDW detector. Such a detector is obtained by setting the free Hamiltonian of the detector to zero, $\hat{H}_d = 0,$ so that no intrinsic energy scale is present. As a consequence, the monopole operator in the interaction picture becomes time-independent, $\hat{m}(\tau) = \hat{m} = \hat{\sigma}_+ + \hat{\sigma}_-$. It is worth noting that, in the following calculations, one could more generally take $\hat{m}$ to be any constant operator acting on the detector Hilbert space satisfying $\hat{m}^2 = \hat{\mathbb{I}}_d$. In the interaction Hamiltonian~\eqref{eq:H_I} $\hat{m}(\tau)$ is replaced by $\hat{m}$. As a consequence, by using the Magnus expansion~\cite{Blanes09}, we can rewrite Eq.~\eqref{eq:U_I} as~\cite{Landulfo2016}
\begin{align}
 \hat{U}_I =  e^{-i \lambda \hat{m} \otimes \hat{\phi}(\Lambda)} e^{-i \mathcal{G} \hat{m}^2}.
\end{align}
where
\begin{align}
    \hat{\phi}(\Lambda) = \int d^4 \mathsf{x} \, \Lambda (\mathsf{x}) \hat{\phi}(\mathsf{x}),
\end{align}
and
\begin{align}
  \mathcal{G} = \frac{\lambda^2}{2} \iint d^4 \mathsf{x} \, d^4 \mathsf{x}' \, \Lambda (\mathsf{x}) \Lambda (\mathsf{x}') \Theta(\tau-\tau') E(\mathsf{x},\mathsf{x}'),
\end{align}
where $ E(\mathsf{x},\mathsf{x}')$ is the unsmeared version of the causal propagator~\eqref{eq:Causalprop}.

In turn, this allows us to write the characteristic function~\eqref{Eq:C(nu)} as
\begin{align}
    C(\nu) = \Tr \left(e^{i \lambda \hat{m} \otimes \hat{\phi}(\Lambda)} e^{i\nu \hat{H}_{\phi}} e^{-i \lambda \hat{m} \otimes \hat{\phi}(\Lambda)} e^{-i \nu \hat{H}_{\phi}}(\hat{\rho}_{d,0} \otimes \hat{\rho}_{\beta})\right).
\end{align}
Moreover, given that 
\begin{align}
    e^{i\nu \hat{H}_{\phi}} e^{-i \lambda \hat{m} \otimes \hat{\phi}(\Lambda)} e^{-i \nu \hat{H}_{\phi}} & = e^{-i \lambda \hat{m}\otimes \int d^4 \mathsf{x} \Lambda(\mathsf{x}) \hat{\phi}(\mathsf{x}(\tau + \nu, \vx))} \nonumber \\
   & = e^{-i\lambda \hat{m}\otimes\hat{\phi}(\Lambda_{\nu})}, 
\end{align}
where $\Lambda_{\nu}(\mathsf{x}) = \Lambda(\mathsf{x}(\tau-\nu, \vx))$, we have
\begin{align}
    C(\nu) = \Tr \left(e^{i \lambda \hat{m} \otimes \hat{\phi}(\Lambda)} e^{-i\lambda \hat{m}\otimes\hat{\phi}(\Lambda_{\nu})}(\hat{\rho}_{d,0} \otimes \hat{\rho}_{\beta})\right).
\end{align}
By applying the Zassenhaus formula, $e^{\mathfrak{a} + \mathfrak{b}} = e^{\mathfrak{a}} e^{\mathfrak{b}} e^{-\frac{1}{2}[\mathfrak{a},\mathfrak{b}]}$, we find that
\begin{align}
    C(\nu) = e^{\frac{i}{2}\lambda^2 E(\Lambda,\Lambda_{\nu})}\Tr \left(e^{i \lambda \hat{m} \otimes \hat{\phi}(\Lambda- \Lambda_{\nu})} (\hat{\rho}_{d,0} \otimes \hat{\rho}_{\beta})\right),\label{eq:Cnugeneral}
\end{align}
where we used $\hat{m}^2= \hat{\mathbb{I}}_d$.  It is worth emphasizing that the above equation does not rely on the field mode decomposition introduced in Eq.~\eqref{eq:modedecomp}, nor have we assumed that the field is in a thermal KMS state. As such, this expression is quite general. This feature makes the present approach particularly suitable for generalizations, for instance, to stationary curved spacetimes.

We can further simplify Eq.~\eqref{eq:Cnugeneral} by using the spectral decomposition $\hat{m} = \sum_j \sigma_j |\sigma_j \rangle \langle \sigma_j|$, i.e.,
\begin{align}
    C(\nu) = e^{\frac{i}{2}\lambda^2 E(\Lambda,\Lambda_{\nu})}\sum_j \rho_{jj,d}\Tr_{\phi}\left(e^{i \lambda \sigma_j \hat{\phi}(\Lambda- \Lambda_{\nu})}\hat{\rho}_{\beta}\right),
\end{align}
where $\rho_{jj,d} = \Tr_d(|\sigma_j \rangle \langle \sigma_j|\hat{\rho}_{d,0})$. Now, using the fact that the KMS state $\rho_{\beta}$ is a quasi-free state~\cite{Kay1991}, we obtain
\begin{align}
    \Tr_{\phi}\left(e^{i \lambda \sigma_j \hat{\phi}(\Lambda- \Lambda_{\nu})}\hat{\rho}_{\beta}\right) = e^{-\frac{1}{2}\lambda^2 \mathcal{W}_{\beta}(\Lambda- \Lambda_{\nu},\Lambda- \Lambda_{\nu})}
\end{align}
where we have used that $\sigma_j^2 = 1$. Given that $\sum_j \rho_{jj,d} = 1$ and
\begin{align}
    \mathcal{W}_{\beta}(\Lambda- \Lambda_{\nu},\Lambda- \Lambda_{\nu}) = & 2\left(\langle(\hat{\phi}(\Lambda))^2\rangle_{\beta}  - \langle\hat{\phi}(\Lambda)\hat{\phi}(\Lambda_{\nu})\rangle_{\beta}\right) \nonumber \\ & + i E(\Lambda,\Lambda_{\nu}),
\end{align}
we arrive at
\begin{align}
     C(\nu) = e^{\lambda^2\left[ \langle\hat{\phi}(\Lambda)\hat{\phi}(\Lambda_{\nu})\rangle_{\beta} - \langle(\hat{\phi}(\Lambda))^2\rangle_{\beta} \right]},
\end{align}
which can still be expressed as
\begin{align}
 C(\nu) =  \exp \Big[\lambda^2 \iint & d^4 \mathsf{x} d^4 \mathsf{x}' \Lambda(\mathsf{x}) \Lambda(\mathsf{x}') \Big(\mathcal{W}_{\beta}(\tau, \tau'+\nu) \nonumber \\ & - \mathcal{W}_{\beta}(\tau, \tau') \Big)\Big]. \label{eq:CnuKMSgapless}
\end{align}
At this point, Eq.~\eqref{eq:CnuKMSgapless} already allows us to recover the detailed fluctuation relation for heat exchange solely from the properties of the thermal Wightman functions, i.e., from the KMS condition. However, we postpone this discussion to the end. Moreover, Eq.~\eqref{eq:CnuKMSgapless} takes the same form for an UDW detector with a delta-switching interaction, for which $\Lambda(x) = \eta \delta(t - t_0) \psi(\textbf{x})$.

By using Eq.~\eqref{eq:KMSw}, the characteristic function takes the form
\begin{align}
 C(\nu) = \exp\Big\{ & \lambda^2 \int \frac{d^3 \textbf{k}}{(2\pi)^32 \omega_{\textbf{k}}} \Big[ \Gamma^+(\omega_{\textbf{k}},\textbf{k}) (e^{i \nu \omega_{\textbf{k}}} - 1) \\ & + \Gamma^-(\omega_{\textbf{k}},\textbf{k})(e^{-i \nu \omega_{\textbf{k}}} - 1) \Big]\Big\}, \nonumber
\end{align}
corresponding exactly to the characteristic function of a bidirectional Poisson process~\cite{Hida2008, Esposito2009}. In this case, the effective rates satisfy the detailed balance condition, given that
\begin{align}
& \Gamma^+(\omega_{\textbf{k}},\textbf{k}) = \frac{e^{\beta \omega_{\textbf{k}}}}{e^{\beta \omega_{\textbf{k}}}-1}  |\tilde{\Lambda}(\omega_{\textbf{k}}, \textbf{k})|^2  \\
& \Gamma^-(\omega_{\textbf{k}},\textbf{k}) =  \frac{1}{e^{\beta \omega_{\textbf{k}}}-1} |\tilde{\Lambda}( \omega_{\textbf{k}}, \textbf{k})|^2.
\end{align}
As a direct consequence, the characteristic function obeys the fluctuation symmetries $C(-\nu + i\beta)= C(\nu)$ and $C(i\beta) = 1$, which implies the detailed fluctuation relation for the exchanged heat of the form $P(Q)/P(-Q) = e^{\beta Q}$ and the Jarzynski equality $\langle e^{-\beta Q}\rangle = 1$. Therefore, in the absence of a detector gap, the energy exchange statistics are fully governed by a thermally balanced bidirectional Poisson process, and the fluctuation relation emerges exactly.

The fact that the detector is gapless plays a crucial role. In the case of a detector with an energy gap, the interaction with the field is spectrally selective: the detector predominantly couples to modes whose frequencies are close to its transition energy. As a consequence, only a restricted portion of the field spectrum effectively contributes to the dynamics, and the resulting energy exchange process depends on both the detector's internal structure and its initial state. In this situation, the fluctuation symmetries, such as $C(i\beta) = 1$, emerge only approximately, reflecting effective fluctuation relations that are not exact due to the finite interaction time and the limited set of modes involved. In contrast, for a gapless detector this spectral selectivity is absent. Since there is no intrinsic energy scale associated with the detector, it couples to all field modes, with weights determined solely by the switching and smearing functions. The absence of spectral filtering allows the characteristic function to directly inherit the symmetry properties of the thermal Wightman functions, which encode the KMS condition. As a consequence, the heat statistics reflects these underlying thermal symmetries, leading to the exact emergence of fluctuation relations.

\section{Conclusions}
\label{Sec:V}
In this work, we investigated the statistics of heat exchange between a quantum field in a thermal KMS state and an Unruh–DeWitt particle detector in Minkowski spacetime. Rather than relying on projective measurements, which are known to be problematic in relativistic quantum field theory, we adopted an interferometric approach in the spirit of Ref.~\cite{Ortega}, allowing for an operational definition of the characteristic function of heat based solely on unitary dynamics. This framework provides a consistent way to access full counting statistics in a relativistic setting, without invoking measurements that may conflict with causality.

Within this approach, at second order in perturbation theory, we derived general expressions for the characteristic function and the corresponding heat distribution. In the vacuum case, the heat statistics is dominated by spontaneous emission processes by the detector, and the corresponding characteristic function resembles that of a unidirectional Poisson process, reflecting the absence of absorption channels. For the field in a KMS state with inverse temperature $\beta$, both emission and absorption processes contribute, and we showed that the characteristic function assumes a form analogous to that of a bidirectional Poisson process in the continuum limit. In this regime, we found that $C(i\beta)\approx 1$, indicating that the fluctuation symmetry is approximately satisfied, with deviations controlled by the detector's initial state and the details of the interaction. We also notice that, when the detector is initially maximally mixed, the fluctuation symmetries are exactly restored at the level of the characteristic function, reflecting the fact that the detector does not introduce any bias between excitation and de-excitation processes and allowing the KMS condition to fully determine the heat statistics.

Then, we restricted our analysis to a massless scalar field and adopted specific choices for the switching and spacetime smearing functions. For a detector with a finite energy gap and finite interaction time, we showed that the resulting detailed fluctuation relations acquire corrections that depend explicitly on the detector's initial state and on the details of the interaction. In particular, deviations from the standard detailed fluctuation relation arise from the spectral selectivity induced by the detector’s internal structure. In this regime, the fluctuation symmetry is only approximate, and its precise form reflects the combined influence of the detector's initial preparation, the switching function, and the thermal field correlations. We further showed that, in the long-time interaction limit, detailed balance emerges at the level of transition probabilities, leading to simplified fluctuation relations that depend only on the initial state of the detector. When the detector is initially prepared in a thermal state with inverse temperature $\beta_D\neq\beta$, the resulting detailed fluctuation relation takes a form analogous to that obtained for heat exchange between systems at different temperatures~\cite{Jarzynski2004}, highlighting the consistency of our framework with known results in quantum thermodynamics.

We also investigated different formulations of the Landauer bound in the context of relativistic quantum field theory. By specializing the algebraic entropy-balance theorem of Jakšić and Pillet to a UDW detector coupled to a field in a KMS state, we obtained the entropic bound $\beta\langle Q\rangle\geq-\Delta S$ without requiring the field state to be represented by a trace-class Gibbs operator. We then evaluated this bound and the complementary full-counting-statistics bound $\beta\langle Q\rangle\geq-\ln C(i\beta)$ using the detector-field dynamics computed to second order in perturbation theory, verifying both inequalities throughout the parameter regimes considered and comparing their dependence on the detector's initial state and interaction time. For the explicitly switched interaction, we further formulated the first-law energy balance in terms of the external switching work and recast the entropic Landauer inequality as a nonequilibrium second-law bound on this work, with the entropy production identified as dissipated switching work. These results show that information-theoretic and thermodynamic constraints on heat exchange extend consistently to detector–field interactions despite the field-theoretic nature and infinitely many degrees of freedom of the environment.

Finally, we analyzed the gapless and delta-switching regimes, and showed that they lead to equivalent descriptions of the heat statistics, allowing for a fully non-perturbative treatment in these cases. In the absence of an intrinsic energy scale, spectral selectivity is lost and the detector effectively couples to all field modes. As a consequence, the characteristic function can be obtained in closed form, yielding a non-perturbative expression entirely determined by the thermal Wightman function. In this regime, the characteristic function takes exactly the same form as that of a bidirectional Poisson process, with rates fixed by the thermal correlations of the field. This leads to an exact fluctuation theorem for heat exchange, in which the full counting statistics directly inherits the symmetry properties encoded in the KMS condition.

Several directions can be pursued to further develop the present work. One natural avenue is to compare our results with the Ramsey-interferometric protocols introduced in Refs.~\cite{Ortega, Bonfill}, which provide a framework for defining work and formulating the first law of quantum field thermodynamics. It is worth emphasizing that, although both approaches rely on Ramsey-type interferometry, the notions of work and heat differ in their operational implementation. In particular, in the work-focused protocols of Refs.~\cite{Ortega}, the setup typically involves only the ancilla and the field, together with a specific controlled unitary between them. In contrast, the heat protocol considered in the present work includes an additional ingredient, namely an UDW detector playing the role of the system of interest, which interacts with the quantum field and thereby exchanges heat with it. This difference suggests that it would be interesting to investigate both protocols within a unified framework, clarifying how work and heat emerge from a common interferometric description. In particular, such a unification may shed light on the role of energy exchange with the field in the formulation of the first law of thermodynamics, where both work and heat contributions must be consistently accounted for, in close connection with the results of Ref.~\cite{Bonfill}.

Another interesting direction concerns the generalization of the present framework to stationary curved spacetimes. For spacetimes admitting a timelike Killing symmetry, such an extension is expected to be relatively straightforward, at least at the level of formal expressions. In particular, since the characteristic functions can be generically expressed in terms of Wightman functions, their structure should retain the same functional form in stationary backgrounds, with the appropriate modifications. This would open the possibility of exploring effects such as the Unruh and Hawking phenomena from the perspective of the present thermodynamic framework. A more challenging extension would be the treatment of genuinely non-stationary spacetimes, where no global notion of time-translation symmetry is available. In such scenarios, the lack of a preferred notion of energy makes the definition of work and heat more subtle, and may require a fully local formulation of the protocol.

%%%%%%%%%%%%%%%%%%%%%%%%%%%%%%%%%%%%%%%%%%%%%%%%%%%%%%%%%%%%%%%
%%%%%%%%%%%%%%%%%%%%%%%%%%%%%%%%%%%%%%%%%%%%%%%%%%%%%%%%%%%%%%%
\begin{acknowledgments}
M. L. W. B. thanks T. Rick Perche for useful discussions on the subject of this manuscript. This work was supported by S\~{a}o Paulo Research Foundation (FAPESP) Grants No.~2025/07325-0 (MLWB) and 2024/00923-6 (AS), and by The Brazilian National Council for Scientific and Technological Development (CNPq), grant
306785/2022-6 (AS).
\end{acknowledgments}

\bibliography{referencias.bib}

@article{Simmons,
  title={A new horizon for quantum information},
  author={Simmons, M.},
  journal={npj Quantum Information},
  volume={15013},
  pages={735},
  year={2015},
  url = {https://doi.org/10.1038/npjqi.2015.13}
}

@article{Adesso,
    author = {Adesso, Gerardo and Franco, Rosario Lo and Parigi, Valentina},
    title = {Foundations of quantum mechanics and their impact on contemporary society},
    journal = {Philos. Trans. A Math. Phys. Eng. Sci.},
    volume = {376},
    number = {2123},
    pages = {20180112},
    year = {2018},
    month = {05},
    doi = {10.1098/rsta.2018.0112},
    url = {https://doi.org/10.1098/rsta.2018.0112},

}

@article{Goold2016a,
doi = {10.1088/1751-8113/49/14/143001},
url = {https://doi.org/10.1088/1751-8113/49/14/143001},
year = {2016},
month = {feb},
volume = {49},
number = {14},
pages = {143001},
author = {Goold, John and Huber, Marcus and Riera, Arnau and Rio, Lídia del and Skrzypczyk, Paul},
title = {The role of quantum information in thermodynamics—a topical review},
journal = {J. Phys. A: Math. Theor.},
}

@article{Xu2022,
  title = {Landauer's principle in qubit-cavity quantum-field-theory interaction in vacuum and thermal states},
  author = {Xu, Hao and Ong, Yen Chin and Yung, Man-Hong},
  journal = {Phys. Rev. A},
  volume = {105},
  issue = {1},
  pages = {012430},
  numpages = {7},
  year = {2022},
  month = {Jan},
  doi = {10.1103/PhysRevA.105.012430},
  url = {https://link.aps.org/doi/10.1103/PhysRevA.105.012430}
}

@article{Witten2018,
  title = {APS Medal for Exceptional Achievement in Research: Invited article on entanglement properties of quantum field theory},
  author = {Witten, Edward},
  journal = {Rev. Mod. Phys.},
  volume = {90},
  issue = {4},
  pages = {045003},
  numpages = {38},
  year = {2018},
  doi = {10.1103/RevModPhys.90.045003},
  url = {https://link.aps.org/doi/10.1103/RevModPhys.90.045003}
}

@article{Dere2003,
author = {Derezi{\'n}ski, J. and Jak{\v s}i{\'c}, V. and Pillet, C.-A.},
title = {Perturbation Theory of W*-Dynamics,  Liouvilleans and KMS-States},
journal = {Reviews in Mathematical Physics},
volume = {15},
number = {05},
pages = {447-489},
year = {2003},
doi = {10.1142/S0129055X03001679},
URL = { https://doi.org/10.1142/S0129055X03001679}
}

@article{Fewster2020,
title = {Quantum Fields and Local Measurements},
journal = {Commun. Math. Phys.},
volume = {378},
pages = {851–889},
year = {2020},
url = {https://doi.org/10.1007/s00220-020-03800-6},
author = {Fewster, C. J. and Verch, R.},
}

@article{Hong2025,
  title = {Work statistics via real-time effective field theory: Application to work extraction from a thermal bath with qubit coupling},
  author = {Hong, Jhh-Jing and Lin, Feng-Li},
  journal = {Phys. Rev. E},
  volume = {112},
  issue = {4},
  pages = {044147},
  numpages = {22},
  year = {2025},
  month = {Oct},
  doi = {10.1103/pt2x-65pq},
  url = {https://link.aps.org/doi/10.1103/pt2x-65pq}
}

@article{Bartolotta,
  title = {Jarzynski Equality for Driven Quantum Field Theories},
  author = {Bartolotta, Anthony and Deffner, Sebastian},
  journal = {Phys. Rev. X},
  volume = {8},
  issue = {1},
  pages = {011033},
  numpages = {20},
  year = {2018},
  month = {Feb},
  publisher = {American Physical Society},
  doi = {10.1103/PhysRevX.8.011033},
  url = {https://link.aps.org/doi/10.1103/PhysRevX.8.011033}
}

@article{Vinjanampathy,
author = {Sai Vinjanampathy and Janet Anders},
title = {Quantum thermodynamics},
journal = {Contemporary Physics},
volume = {57},
number = {4},
pages = {545--579},
year = {2016},
doi = {10.1080/00107514.2016.1201896},
URL = {https://doi.org/10.1080/00107514.2016.1201896}
}

@article{Alsing2003,
  title = {Teleportation with a Uniformly Accelerated Partner},
  author = {Alsing, Paul M. and Milburn, G. J.},
  journal = {Phys. Rev. Lett.},
  volume = {91},
  issue = {18},
  pages = {180404},
  numpages = {4},
  year = {2003},
  month = {Oct},
  doi = {10.1103/PhysRevLett.91.180404},
  url = {https://link.aps.org/doi/10.1103/PhysRevLett.91.180404}
}

@article{Evans,
  title = {Probability of second law violations in shearing steady states},
  author = {Evans, Denis J. and Cohen, E. G. D. and Morriss, G. P.},
  journal = {Phys. Rev. Lett.},
  volume = {71},
  issue = {15},
  pages = {2401--2404},
  numpages = {0},
  year = {1993},
  month = {Oct},
  doi = {10.1103/PhysRevLett.71.2401},
  url = {https://link.aps.org/doi/10.1103/PhysRevLett.71.2401}
}

@ARTICLE{Crooks1998,
       author = {{Crooks}, Gavin E.},
        title = {Nonequilibrium Measurements of Free Energy Differences for Microscopically Reversible Markovian Systems},
      journal = {Journal of Statistical Physics},
         year = {1998},
        month = {mar},
       volume = {90},
       number = {5-6},
        pages = {1481-1487},
          doi = {10.1023/A:1023208217925}
}

@article{Jarzynski2004,
  title = {Classical and Quantum Fluctuation Theorems for Heat Exchange},
  author = {Jarzynski, Christopher and W\'ojcik, Daniel K.},
  journal = {Phys. Rev. Lett.},
  volume = {92},
  issue = {23},
  pages = {230602},
  numpages = {4},
  year = {2004},
  month = {Jun},
  doi = {10.1103/PhysRevLett.92.230602},
  url = {https://link.aps.org/doi/10.1103/PhysRevLett.92.230602}
}

@ARTICLE{Papageorgiou,
       author = {{Papageorgiou}, Maria and {Fraser}, Doreen},
        title = {Eliminating the 'Impossible': Recent Progress on Local Measurement Theory for Quantum Field Theory},
      journal = {Foundations of Physics},
         year = {2024},
        month = {jun},
       volume = {54},
       number = {3},
          eid = {26},
        pages = {26},
          doi = {10.1007/s10701-024-00756-8}
}

@article{Papadatos,
  title = {Relativistic quantum thermodynamics of moving systems},
  author = {Papadatos, Nikolaos and Anastopoulos, Charis},
  journal = {Phys. Rev. D},
  volume = {102},
  issue = {8},
  pages = {085005},
  numpages = {15},
  year = {2020},
  month = {Oct},
  publisher = {American Physical Society},
  doi = {10.1103/PhysRevD.102.085005},
  url = {https://link.aps.org/doi/10.1103/PhysRevD.102.085005}
}

@ARTICLE{Good2020,
       author = {{Good}, Michael and {Ju{\'a}rez-Aubry}, Benito A. and {Moustos}, Dimitris and {Temirkhan}, Maksat},
        title = {Unruh-like effects: effective temperatures along stationary worldlines},
      journal = {Journal of High Energy Physics},
         year = {2020},
        month = {jun},
       volume = {2020},
       number = {6},
          eid = {59},
        pages = {59},
          doi = {10.1007/JHEP06(2020)059}
}

@article{Seifert2012,
doi = {10.1088/0034-4885/75/12/126001},
url = {https://doi.org/10.1088/0034-4885/75/12/126001},
year = {2012},
month = {nov},
volume = {75},
number = {12},
pages = {126001},
author = {Seifert, Udo},
title = {Stochastic thermodynamics, fluctuation theorems and molecular machines},
journal = {Reports on Progress in Physics}
}

@article{Valentini,
title = {Non-local correlations in quantum electrodynamics},
journal = {Phys. Lett. A},
volume = {153},
number = {6},
pages = {321-325},
year = {1991},
issn = {0375-9601},
doi = {https://doi.org/10.1016/0375-9601(91)90952-5},
url = {https://www.sciencedirect.com/science/article/pii/0375960191909525},
author = {Valentini, A.}
}

@article{Pozas2015,
  title = {Harvesting correlations from the quantum vacuum},
  author = {Pozas-Kerstjens, Alejandro and Mart\'{\i}n-Mart\'{\i}nez, Eduardo},
  journal = {Phys. Rev. D},
  volume = {92},
  issue = {6},
  pages = {064042},
  numpages = {18},
  year = {2015},
  month = {Sep},
  publisher = {American Physical Society},
  doi = {10.1103/PhysRevD.92.064042},
  url = {https://link.aps.org/doi/10.1103/PhysRevD.92.064042}
}

@article{Resnik,
       author = {Reznik, Benni},
        title = {Entanglement from the Vacuum},
      journal = {Foundations of Physics},
         year = {2003},
       volume = {33},
       number = {1},
        pages = {167-176},
          doi = {10.1023/A:1022875910744},
}

@article{Perche2024,
  title = {Fully relativistic entanglement harvesting},
  author = {Perche, T. Rick and Polo-G\'omez, Jos\'e and Torres, Bruno de S. L. and Mart\'{\i}n-Mart\'{\i}nez, Eduardo},
  journal = {Phys. Rev. D},
  volume = {109},
  issue = {4},
  pages = {045018},
  numpages = {17},
  year = {2024},
  month = {Feb},
  publisher = {American Physical Society},
  doi = {10.1103/PhysRevD.109.045018},
  url = {https://link.aps.org/doi/10.1103/PhysRevD.109.045018}
}

@article{Jonsson2015,
  title = {Information Transmission Without Energy Exchange},
  author = {Jonsson, Robert H. and Mart\'{\i}n-Mart\'{\i}nez, Eduardo and Kempf, Achim},
  journal = {Phys. Rev. Lett.},
  volume = {114},
  issue = {11},
  pages = {110505},
  numpages = {5},
  year = {2015},
  month = {Mar},
  publisher = {American Physical Society},
  doi = {10.1103/PhysRevLett.114.110505},
  url = {https://link.aps.org/doi/10.1103/PhysRevLett.114.110505}
}

@article{Hotta2008,
  title = {Quantum measurement information as a key to energy extraction from local vacuums},
  author = {Hotta, Masahiro},
  journal = {Phys. Rev. D},
  volume = {78},
  issue = {4},
  pages = {045006},
  numpages = {9},
  year = {2008},
  month = {Aug},
  doi = {10.1103/PhysRevD.78.045006},
  url = {https://link.aps.org/doi/10.1103/PhysRevD.78.045006}
}

@article{Preskill,
author = {Preskill, J.},
title = {Quantum information and physics: Some future directions},
journal = {Journal of Modern Optics},
volume = {47},
number = {2-3},
pages = {127--137},
year = {2000},
doi = {10.1080/09500340008244031},
URL = {https://www.tandfonline.com/doi/abs/10.1080/09500340008244031}
}

@article{Jaynes1957a,
  title = {Information Theory and Statistical Mechanics},
  author = {Jaynes, E. T.},
  journal = {Phys. Rev.},
  volume = {106},
  issue = {4},
  pages = {620--630},
  numpages = {0},
  year = {1957},
  month = {May},
  doi = {10.1103/PhysRev.106.620},
  url = {https://link.aps.org/doi/10.1103/PhysRev.106.620}
}

@article{Jaynes1957b,
  title = {Information Theory and Statistical Mechanics. II},
  author = {Jaynes, E. T.},
  journal = {Phys. Rev.},
  volume = {108},
  issue = {2},
  pages = {171--190},
  numpages = {0},
  year = {1957},
  month = {Oct},
  doi = {10.1103/PhysRev.108.171},
  url = {https://link.aps.org/doi/10.1103/PhysRev.108.171}
}

@article{Landauer,
  author={Landauer, R.},
  journal={IBM Journal of Research and Development}, 
  title={Irreversibility and heat generation in the computing process}, 
  year={2000},
  volume={44},
  number={1.2},
  pages={261-269},
  doi={10.1147/rd.441.0261}
  }

@article{Bennett,
    author = "Bennett, Charles H.",
    title = "{The thermodynamics of computation{\textemdash}a review}",
    doi = "10.1007/BF02084158",
    journal = "Int. J. Theor. Phys.",
    volume = "21",
    number = "12",
    pages = "905--940",
    year = "1982"
}

@article{Plenio,
author = {M. B. Plenio and V. Vitelli},
title = {The physics of forgetting: Landauer's erasure principle and information theory},
journal = {Contemporary Physics},
volume = {42},
number = {1},
pages = {25--60},
year = {2001},
publisher = {Taylor \& Francis},
doi = {10.1080/00107510010018916},
URL = {https://doi.org/10.1080/00107510010018916},
}

@book{Hida2008,
  title={Lectures on White Noise Functionals},
  author={Hida, T. and Si Si},
  year={2008},
  publisher={World Scientific Publishing}
}

@book{Alicki,
  title={Introduction to quantum thermodynamics: History and prospects, in
Thermodynamics in the Quantum Regime},
  author={R. Alicki and R. Kosloff},
  year={2018},
  publisher={Springer International Publishing}
}

@article{Landi2021,
  title = {Irreversible entropy production: From classical to quantum},
  author = {Landi, Gabriel T. and Paternostro, Mauro},
  journal = {Rev. Mod. Phys.},
  volume = {93},
  issue = {3},
  pages = {035008},
  numpages = {58},
  year = {2021},
  month = {Sep},
  doi = {10.1103/RevModPhys.93.035008},
  url = {https://link.aps.org/doi/10.1103/RevModPhys.93.035008}
}

@article{Unruh1976,
  author = {Unruh, W. G.},
  title = {Notes on black-hole evaporation},
  journal = {Phys. Rev. D},
  volume = {14},
  pages = {870},
  year = {1976},
  url = {https://doi.org/10.1103/PhysRevD.14.870}
}

@article{Costa2026,
  title = {Work distribution of quantum fields in static curved spacetimes},
  author = {Costa, Rafael L. S. and Basso, Marcos L. W. and Maziero, Jonas and C\'eleri, Lucas C.},
  journal = {Phys. Rev. D},
  volume = {113},
  issue = {2},
  pages = {025010},
  numpages = {10},
  year = {2026},
  month = {Jan},
  publisher = {American Physical Society},
  doi = {10.1103/d26w-cj7z},
  url = {https://link.aps.org/doi/10.1103/d26w-cj7z}
}

@article{Ramezani,
  title = {Fluctuation relation for heat exchange in Markovian open quantum systems},
  author = {Ramezani, M. and Golshani, M. and Rezakhani, A. T.},
  journal = {Phys. Rev. E},
  volume = {97},
  issue = {4},
  pages = {042101},
  numpages = {5},
  year = {2018},
  month = {Apr},
  publisher = {American Physical Society},
  doi = {10.1103/PhysRevE.97.042101},
  url = {https://link.aps.org/doi/10.1103/PhysRevE.97.042101}
}

@article{Campisi2011,
  author = {Campisi, M. and Talkner, P. and H\"anggi, P.},
  title = {Colloquium: Quantum fluctuation relations: Foundations and applications},
  journal = {Rev. Mod. Phys.},
  volume = {83},
  pages = {771},
  year = {2011},
  url = {https://doi.org/10.1103/RevModPhys.83.771}
}

@article{Chaves2025,
  title = {A Friendly Guide to Exorcising Maxwell's Demon},
  author = {de Oliveira Junior, A. and Brask, Jonatan Bohr and Chaves, Rafael},
  journal = {PRX Quantum},
  volume = {6},
  issue = {3},
  pages = {030201},
  numpages = {47},
  year = {2025},
  month = {Aug},
  publisher = {American Physical Society},
  doi = {10.1103/phkv-wrsd},
  url = {https://link.aps.org/doi/10.1103/phkv-wrsd}
}

@article{Sorkin1993,
  author = {Sorkin, R. D.},
  title = {Impossible measurements on quantum fields},
  journal = {Directions in General Relativity},
  volume = {2},
  pages = {293},
  year = {1993},
  publisher = {Cambridge University Press},
  url = {https://arxiv.org/abs/gr-qc/9302018v2},
}

@article{Bostelmann,
  title = {Impossible measurements require impossible apparatus},
  author = {Bostelmann, Henning and Fewster, Christopher J. and Ruep, Maximilian H.},
  journal = {Phys. Rev. D},
  volume = {103},
  pages = {025017},
  year = {2021},
  doi = {10.1103/PhysRevD.103.025017},
  url = {https://link.aps.org/doi/10.1103/PhysRevD.103.025017}
}

@article{Borsten,
  title = {Impossible measurements revisited},
  author = {Borsten, L. and Jubb, I. and Kells, G.},
  journal = {Phys. Rev. D},
  volume = {104},
  issue = {2},
  pages = {025012},
  numpages = {8},
  year = {2021},
  month = {Jul},
  publisher = {American Physical Society},
  doi = {10.1103/PhysRevD.104.025012},
  url = {https://link.aps.org/doi/10.1103/PhysRevD.104.025012}
}

@article{Bonfill,
  title = {First law of quantum field thermodynamics},
  author = {Teixid\'o-Bonfill, Adam and Ortega, Alvaro and Mart\'{\i}n-Mart\'{\i}nez, Eduardo},
  journal = {Phys. Rev. A},
  volume = {102},
  issue = {5},
  pages = {052219},
  numpages = {10},
  year = {2020},
  month = {Nov},
  publisher = {American Physical Society},
  doi = {10.1103/PhysRevA.102.052219},
  url = {https://link.aps.org/doi/10.1103/PhysRevA.102.052219}
}

@article{Ortega,
  title = {Work Distributions on Quantum Fields},
  author = {Ortega, Alvaro and McKay, Emma and Alhambra, \'Alvaro M. and Mart\'{\i}n-Mart\'{\i}nez, Eduardo},
  journal = {Phys. Rev. Lett.},
  volume = {122},
  issue = {24},
  pages = {240604},
  numpages = {6},
  year = {2019},
  month = {Jun},
  publisher = {American Physical Society},
  doi = {10.1103/PhysRevLett.122.240604},
  url = {https://link.aps.org/doi/10.1103/PhysRevLett.122.240604}
}

@article{Aubry,
  title = {Asymptotic states for stationary Unruh-DeWitt detectors},
  author = {Ju\'arez-Aubry, Benito A. and Moustos, Dimitris},
  journal = {Phys. Rev. D},
  volume = {100},
  issue = {2},
  pages = {025018},
  numpages = {12},
  year = {2019},
  month = {Jul},
  publisher = {American Physical Society},
  doi = {10.1103/PhysRevD.100.025018},
  url = {https://link.aps.org/doi/10.1103/PhysRevD.100.025018}
}

@article{Garay,
  title = {Thermalization of particle detectors: The Unruh effect and its reverse},
  author = {Garay, Luis J. and Mart\'{\i}n-Mart\'{\i}nez, Eduardo and de Ram\'on, Jos\'e},
  journal = {Phys. Rev. D},
  volume = {94},
  issue = {10},
  pages = {104048},
  numpages = {11},
  year = {2016},
  month = {Nov},
  publisher = {American Physical Society},
  doi = {10.1103/PhysRevD.94.104048},
  url = {https://link.aps.org/doi/10.1103/PhysRevD.94.104048}
}

@article{Perche2021,
  title = {General features of the thermalization of particle detectors and the Unruh effect},
  author = {Perche, T. Rick},
  journal = {Phys. Rev. D},
  volume = {104},
  issue = {6},
  pages = {065001},
  numpages = {14},
  year = {2021},
  month = {Sep},
  publisher = {American Physical Society},
  doi = {10.1103/PhysRevD.104.065001},
  url = {https://link.aps.org/doi/10.1103/PhysRevD.104.065001}
}

@article{Tjoa2023,
  title = {Nonperturbative simple-generated interactions with a quantum field for arbitrary Gaussian states},
  author = {Tjoa, Erickson},
  journal = {Phys. Rev. D},
  volume = {108},
  issue = {4},
  pages = {045003},
  numpages = {18},
  year = {2023},
  month = {Aug},
  publisher = {American Physical Society},
  doi = {10.1103/PhysRevD.108.045003},
  url = {https://link.aps.org/doi/10.1103/PhysRevD.108.045003}
}

@article{Shah2025,
  title = {Relativistic QFT description for the interaction of a spin with a magnetic field},
  author = {Shah, Ruhi and Mart\'{\i}n-Mart\'{\i}nez, Eduardo and Perche, T. Rick},
  journal = {Phys. Rev. D},
  volume = {111},
  issue = {4},
  pages = {044075},
  numpages = {20},
  year = {2025},
  month = {Feb},
  doi = {10.1103/PhysRevD.111.044075},
  url = {https://link.aps.org/doi/10.1103/PhysRevD.111.044075}
}

@article{Polo,
  title = {A detector-based measurement theory for quantum field theory},
  author = {Polo-G\'omez, Jos\'e and Garay, Luis J. and Mart\'{\i}n-Mart\'{\i}nez, Eduardo},
  journal = {Phys. Rev. D},
  volume = {105},
  issue = {6},
  pages = {065003},
  numpages = {29},
  year = {2022},
  month = {Mar},
  publisher = {American Physical Society},
  doi = {10.1103/PhysRevD.105.065003},
  url = {https://link.aps.org/doi/10.1103/PhysRevD.105.065003}
}

@article{Fewster,
  title = {Waiting for Unruh},
  author = {Fewster, C. J. and Juárez-Aubry, B. A. and Louko, J.},
  journal = {Class. Quantum Grav.},
  volume = {33},
  pages = {165003},
  year = {2016},
  url = {https://iopscience.iop.org/article/10.1088/0264-9381/33/16/165003}
}

@Article{Anastopoulos2022,
AUTHOR = {Anastopoulos, Charis and Savvidou, Ntina},
TITLE = {Quantum Information in Relativity: The Challenge of QFT Measurements},
JOURNAL = {Entropy},
VOLUME = {24},
YEAR = {2022},
NUMBER = {1},
ARTICLE-NUMBER = {4},
URL = {https://www.mdpi.com/1099-4300/24/1/4}
}

@article{Iso2011,
title = {Non-equilibrium fluctuations of black hole horizons and the generalized second law},
journal = {Phys. Lett. B},
volume = {705},
number = {1},
pages = {152-156},
year = {2011},
issn = {0370-2693},
doi = {https://doi.org/10.10/16/j.physletb.2011.09.114},
url = {https://www.sciencedirect.com/science/article/pii/S0370269311012147},
author = {Satoshi Iso and Susumu Okazawa and Sen Zhang},
}

@article{Reeb2014,
doi = {10.1088/1367-2630/16/10/103011},
url = {https://doi.org/10.1088/1367-2630/16/10/103011},
year = {2014},
month = {oct},
publisher = {IOP Publishing},
volume = {16},
number = {10},
pages = {103011},
author = {Reeb, David and Wolf, Michael M},
title = {An improved Landauer principle with finite-size corrections},
journal = {New J. Phys.}}

@article{Campbell2017,
  title = {Nonequilibrium quantum bounds to Landauer's principle: Tightness and effectiveness},
  author = {Campbell, Steve and Guarnieri, Giacomo and Paternostro, Mauro and Vacchini, Bassano},
  journal = {Phys. Rev. A},
  volume = {96},
  issue = {4},
  pages = {042109},
  numpages = {7},
  year = {2017},
  month = {Oct},
  publisher = {American Physical Society},
  doi = {10.1103/PhysRevA.96.042109},
  url = {https://link.aps.org/doi/10.1103/PhysRevA.96.042109}
}

@article{Guarnieri2017,
doi = {10.1088/1367-2630/aa8cf1},
url = {https://doi.org/10.1088/1367-2630/aa8cf1},
year = {2017},
month = {nov},
publisher = {IOP Publishing},
volume = {19},
number = {10},
pages = {103038},
author = {Guarnieri, Giacomo and Campbell, Steve and Goold, John and Pigeon, Simon and Vacchini, Bassano and Paternostro, Mauro},
title = {Full counting statistics approach to the quantum non-equilibrium Landauer bound},
journal = {New J. Phys.}
}

@article{Hashimoto,
  title = {Lower bounds for the mean dissipated heat in an open quantum system},
  author = {Hashimoto, Kazunari and Vacchini, Bassano and Uchiyama, Chikako},
  journal = {Phys. Rev. A},
  volume = {101},
  issue = {5},
  pages = {052114},
  numpages = {8},
  year = {2020},
  month = {May},
  doi = {10.1103/PhysRevA.101.052114},
  url = {https://link.aps.org/doi/10.1103/PhysRevA.101.052114}
}

@article{Peterson,
    author = {Peterson, J. P. S. and Sarthour, R. S. and Souza, A. M. and Oliveira, I. S. and Goold, J. and Modi, K. and Soares-Pinto, D. O. and Céleri, L. C.},
    title = {Experimental demonstration of information to energy conversion in a quantum system at the Landauer limit},
    journal = {Proc. R. Soc. A},
    volume = {472},
    number = {2188},
    pages = {20150813},
    year = {2016},
    month = {04},
    doi = {10.1098/rspa.2015.0813},
    url = {https://doi.org/10.1098/rspa.2015.0813},

}

@article{Bekenstein74,
  title = {Generalized second law of thermodynamics in black-hole physics},
  author = {Bekenstein, Jacob D.},
  journal = {Phys. Rev. D},
  volume = {9},
  issue = {12},
  pages = {3292--3300},
  numpages = {0},
  year = {1974},
  month = {Jun},
  publisher = {American Physical Society},
  doi = {10.1103/PhysRevD.9.3292},
  url = {https://link.aps.org/doi/10.1103/PhysRevD.9.3292}
}

@article{Landulfo2016,
  title = {Nonperturbative approach to relativistic quantum communication channels},
  author = {Landulfo, Andr\'e G. S.},
  journal = {Phys. Rev. D},
  volume = {93},
  issue = {10},
  pages = {104019},
  numpages = {13},
  year = {2016},
  month = {May},
  publisher = {American Physical Society},
  doi = {10.1103/PhysRevD.93.104019},
  url = {https://link.aps.org/doi/10.1103/PhysRevD.93.104019}
}

@article{Martinez20,
  title = {General relativistic quantum optics: Finite-size particle detector models in curved spacetimes},
  author = {Mart\'{\i}n-Mart\'{\i}nez, Eduardo and Perche, T. Rick and de S. L. Torres, Bruno},
  journal = {Phys. Rev. D},
  volume = {101},
  issue = {4},
  pages = {045017},
  numpages = {10},
  year = {2020},
  month = {Feb},
  publisher = {American Physical Society},
  doi = {10.1103/PhysRevD.101.045017},
  url = {https://link.aps.org/doi/10.1103/PhysRevD.101.045017}
}

@article{Martinez21,
  title = {Broken covariance of particle detector models in relativistic quantum information},
  author = {Mart\'{\i}n-Mart\'{\i}nez, Eduardo and Perche, T. Rick and Torres, Bruno de S. L.},
  journal = {Phys. Rev. D},
  volume = {103},
  issue = {2},
  pages = {025007},
  numpages = {14},
  year = {2021},
  month = {Jan},
  publisher = {American Physical Society},
  doi = {10.1103/PhysRevD.103.025007},
  url = {https://link.aps.org/doi/10.1103/PhysRevD.103.025007}
}

@article{Perche2023,
  title = {Role of quantum degrees of freedom of relativistic fields in quantum information protocols},
  author = {Perche, T. Rick and Mart\'{\i}n-Mart\'{\i}nez, Eduardo},
  journal = {Phys. Rev. A},
  volume = {107},
  issue = {4},
  pages = {042612},
  numpages = {20},
  year = {2023},
  month = {Apr},
  doi = {10.1103/PhysRevA.107.042612},
  url = {https://link.aps.org/doi/10.1103/PhysRevA.107.042612}
}

@article{Blanes09,
title = {The Magnus expansion and some of its applications},
journal = {Phys. Rep.},
volume = {470},
number = {5},
pages = {151-238},
year = {2009},
issn = {0370-1573},
doi = {https://doi.org/10.1016/j.physrep.2008.11.001},
url = {https://www.sciencedirect.com/science/article/pii/S0370157308004092},
author = {S. Blanes and F. Casas and J.A. Oteo and J. Ros},
}

@article{Esposito2010,
doi = {10.1088/1367-2630/12/1/013013},
url = {https://doi.org/10.1088/1367-2630/12/1/013013},
year = {2010},
month = {jan},
volume = {12},
number = {1},
pages = {013013},
author = {Esposito, Massimiliano and Lindenberg, Katja and Van den Broeck, Christian},
title = {Entropy production as correlation between system and reservoir},
journal = {New J. Phys.}
}

@article{Goold2014,
  title = {Measuring the heat exchange of a quantum process},
  author = {Goold, John and Poschinger, Ulrich and Modi, Kavan},
  journal = {Phys. Rev. E},
  volume = {90},
  issue = {2},
  pages = {020101},
  numpages = {4},
  year = {2014},
  month = {Aug},
  doi = {10.1103/PhysRevE.90.020101},
  url = {https://link.aps.org/doi/10.1103/PhysRevE.90.020101}
}

@article{Schlicht04,
doi = {10.1088/0264-9381/21/19/011},
url = {https://dx.doi.org/10.1088/0264-9381/21/19/011},
year = {2004},
month = {sep},
volume = {21},
number = {19},
pages = {4647},
author = {Sebastian Schlicht},
title = {Considerations on the Unruh effect: causality and regularization},
journal = {Class. Quantum Grav},
}

@article{Louko06,
doi = {10.1088/0264-9381/23/22/015},
url = {https://dx.doi.org/10.1088/0264-9381/23/22/015},
year = {2006},
month = {oct},
volume = {23},
number = {22},
pages = {6321},
author = {Louko, Jorma and Satz, Alejandro},
title = {How often does the Unruh–DeWitt detector click? Regularization by a spatial profile},
journal = {Class. Quantum Grav.},
}

@article{Taranto,
  title = {Emergence of a fluctuation relation for heat in nonequilibrium Landauer processes},
  author = {Taranto, Philip and Modi, Kavan and Pollock, Felix A.},
  journal = {Phys. Rev. E},
  volume = {97},
  issue = {5},
  pages = {052111},
  numpages = {8},
  year = {2018},
  month = {May},
  doi = {10.1103/PhysRevE.97.052111},
  url = {https://link.aps.org/doi/10.1103/PhysRevE.97.052111}
}

@article{Hilt,
  title = {Landauer's principle in the quantum regime},
  author = {Hilt, Stefanie and Shabbir, Saroosh and Anders, Janet and Lutz, Eric},
  journal = {Phys. Rev. E},
  volume = {83},
  issue = {3},
  pages = {030102},
  numpages = {4},
  year = {2011},
  month = {Mar},
  publisher = {American Physical Society},
  doi = {10.1103/PhysRevE.83.030102},
  url = {https://link.aps.org/doi/10.1103/PhysRevE.83.030102}
}

@article{Perche2024a,
  title = {Closed-form expressions for smeared bidistributions of a massless scalar field: Nonperturbative and asymptotic results in relativistic quantum information},
  author = {Perche, T. Rick},
  journal = {Phys. Rev. D},
  volume = {110},
  issue = {2},
  pages = {025013},
  numpages = {21},
  year = {2024},
  month = {Jul},
  doi = {10.1103/PhysRevD.110.025013},
  url = {https://link.aps.org/doi/10.1103/PhysRevD.110.025013}
}

@article{Taranto2023,
  title = {Landauer Versus Nernst: What is the True Cost of Cooling a Quantum System?},
  author = {Taranto, Philip and Bakhshinezhad, Faraj and Bluhm, Andreas and Silva, Ralph and Friis, Nicolai and Lock, Maximilian P.E. and Vitagliano, Giuseppe and Binder, Felix C. and Debarba, Tiago and Schwarzhans, Emanuel and Clivaz, Fabien and Huber, Marcus},
  journal = {PRX Quantum},
  volume = {4},
  issue = {1},
  pages = {010332},
  numpages = {61},
  year = {2023},
  month = {Mar},
  doi = {10.1103/PRXQuantum.4.010332},
  url = {https://link.aps.org/doi/10.1103/PRXQuantum.4.010332}
}

@article{Louko08,
doi = {10.1088/0264-9381/25/5/055012},
url = {https://dx.doi.org/10.1088/0264-9381/25/5/055012},
year = {2008},
month = {feb},
volume = {25},
number = {5},
pages = {055012},
author = {Louko, Jorma and Satz, Alejandro},
title = {Transition rate of the Unruh–DeWitt detector in curved spacetime},
journal = {Class. Quantum Grav.},
}

@article{Crooks1999,
  author = {Gavin E. Crooks},
  title = {Entropy production fluctuation theorem and the nonequilibrium work relation for free energy differences},
  journal = {Phys. Rev. E},
  volume = {60},
  number = {3},
  pages = {2721--2726},
  year = {1999},
  doi = {10.1103/PhysRevE.60.2721}
}

@article{Jarzynski1997,
  author = {Christopher Jarzynski},
  title = {Nonequilibrium equality for free energy differences},
  journal = {Phys. Rev. Lett.},
  volume = {78},
  number = {14},
  pages = {2690--2693},
  year = {1997},
  doi = {10.1103/PhysRevLett.78.2690}
}

@article{Goold2015,
  title = {Nonequilibrium Quantum Landauer Principle},
  author = {Goold, John and Paternostro, Mauro and Modi, Kavan},
  journal = {Phys. Rev. Lett.},
  volume = {114},
  issue = {6},
  pages = {060602},
  numpages = {5},
  year = {2015},
  month = {Feb},
  doi = {10.1103/PhysRevLett.114.060602},
  url = {https://link.aps.org/doi/10.1103/PhysRevLett.114.060602}
}

@article{Jarzynski11,
   author = {Jarzynski, Christopher},
   title = {Equalities and Inequalities: Irreversibility and the Second Law of Thermodynamics at the Nanoscale}, 
   journal= {Annual Review of Condensed Matter Physics},
   year = {2011},
   volume = {2},
   pages = {329-351},
   doi = {https://doi.org/10.1146/annurev-conmatphys-062910-140506},
   url = {https://www.annualreviews.org/content/journals/10.1146/annurev-conmatphys-062910-140506},
}

@article{Dorner2013,
  author = {Ralph Dorner and Stephen R. Clark and Libby Heaney and Raymond Fazio and Tobias Calarco and Dieter Jaksch},
  title = {Extracting quantum work statistics and fluctuation theorems by single-qubit interferometry},
  journal = {Phys. Rev. Lett.},
  volume = {110},
  number = {23},
  pages = {230601},
  year = {2013},
  doi = {10.1103/PhysRevLett.110.230601}
}

@article{Mazzola2013,
  author = {Luca Mazzola and G. De Chiara and Mauro Paternostro},
  title = {Measuring the characteristic function of the work distribution},
  journal = {Phys. Rev. Lett.},
  volume = {110},
  number = {23},
  pages = {230602},
  year = {2013},
  doi = {10.1103/PhysRevLett.110.230602}
}

@article{Esposito2009,
  author = {Massimiliano Esposito and Upendra Harbola and Shaul Mukamel},
  title = {Nonequilibrium fluctuations, fluctuation theorems, and counting statistics in quantum systems},
  journal = {Rev. Mod. Phys.},
  volume = {81},
  number = {4},
  pages = {1665--1702},
  year = {2009},
  doi = {10.1103/RevModPhys.81.1665},
  url = {https://doi.org/10.1103/RevModPhys.81.1665}
}

@article{Peres2004,
  author = {Asher Peres and Daniel R. Terno},
  title = {Quantum information and relativity theory},
  journal = {Reviews of Modern Physics},
  volume = {76},
  number = {1},
  pages = {93--123},
  year = {2004},
  doi = {10.1103/RevModPhys.76.93}
}

@inproceedings{DeWitt1979,
  author = {Bryce S. DeWitt},
  title = {Quantum gravity: the new synthesis},
  booktitle = {General Relativity: An Einstein Centenary Survey},
  editor = {S. W. Hawking and W. Israel},
  pages = {680--745},
  year = {1979},
  publisher = {Cambridge University Press}
}

@article{Takagi1986,
  author = {Shuji Takagi},
  title = {Vacuum noise and stress induced by uniform acceleration: Hawking-Unruh effect in Rindler manifold of arbitrary dimension},
  journal = {Progress of Theoretical Physics Supplement},
  volume = {88},
  pages = {1--142},
  year = {1986},
  doi = {10.1143/PTPS.88.1}
}

@article{Kay1991,
title = {Theorems on the uniqueness and thermal properties of stationary, nonsingular, quasifree states on spacetimes with a bifurcate killing horizon},
journal = {Phys. Rep.},
volume = {207},
number = {2},
pages = {49-136},
year = {1991},
issn = {0370-1573},
doi = {https://doi.org/10.1016/0370-1573(91)90015-E},
url = {https://www.sciencedirect.com/science/article/pii/037015739190015E},
author = {Bernard S. Kay and Robert M. Wald},
}

@article{Satz2007,
  author = {Alexander Satz},
  title = {Then again, how often does the Unruh-DeWitt detector click if we switch it carefully?},
  journal = {Class. Quantum Grav.},
  volume = {24},
  number = {7},
  pages = {1719--1732},
  year = {2007},
  doi = {10.1088/0264-9381/24/7/003}
}

@book{Haag1996,
  author = {Rudolf Haag},
  title = {Local Quantum Physics: Fields, Particles, Algebras},
  publisher = {Springer},
  year = {1996},
  edition = {2nd}
}

@article{Ramon23,
  title = {Causality and signalling in noncompact detector-field interactions},
  author = {de Ram\'on, Jos\'e and Papageorgiou, Maria and Mart\'{\i}n-Mart\'{\i}nez, Eduardo},
  journal = {Phys. Rev. D},
  volume = {108},
  issue = {4},
  pages = {045015},
  numpages = {26},
  year = {2023},
  month = {Aug},
  publisher = {American Physical Society},
  doi = {10.1103/PhysRevD.108.045015},
  url = {https://link.aps.org/doi/10.1103/PhysRevD.108.045015}
}

@article{Ramon21,
  title = {Relativistic causality in particle detector models: Faster-than-light signaling and impossible measurements},
  author = {de Ram\'on, Jos\'e and Papageorgiou, Maria and Mart\'{\i}n-Mart\'{\i}nez, Eduardo},
  journal = {Phys. Rev. D},
  volume = {103},
  issue = {8},
  pages = {085002},
  numpages = {13},
  year = {2021},
  month = {Apr},
  publisher = {American Physical Society},
  doi = {10.1103/PhysRevD.103.085002},
  url = {https://link.aps.org/doi/10.1103/PhysRevD.103.085002}
}

@article{Cai25,
title = {General relativistic fluctuation theorems},
journal = {Phys. Lett. B},
volume = {860},
pages = {139220},
year = {2025},
issn = {0370-2693},
doi = {https://doi.org/10.1016/j.physletb.2024.139220},
url = {https://www.sciencedirect.com/science/article/pii/S0370269324007780},
author = {Yifan Cai and Tao Wang and Liu Zhao}
}

@article{Tjoa22,
  title = {Fermi two-atom problem: Nonperturbative approach via relativistic quantum information and algebraic quantum field theory},
  author = {Tjoa, Erickson},
  journal = {Phys. Rev. D},
  volume = {106},
  issue = {4},
  pages = {045012},
  numpages = {13},
  year = {2022},
  month = {Aug},
  publisher = {American Physical Society},
  doi = {10.1103/PhysRevD.106.045012},
  url = {https://link.aps.org/doi/10.1103/PhysRevD.106.045012}
}

@article{Tjoa2022a,
  title = {Quantum teleportation with relativistic communication from first principles},
  author = {Tjoa, Erickson},
  journal = {Phys. Rev. A},
  volume = {106},
  issue = {3},
  pages = {032432},
  numpages = {14},
  year = {2022},
  month = {Sep},
  doi = {10.1103/PhysRevA.106.032432},
  url = {https://link.aps.org/doi/10.1103/PhysRevA.106.032432}
}

@article{Hu2012,
doi = {10.1088/0264-9381/29/22/224005},
url = {https://doi.org/10.1088/0264-9381/29/22/224005},
year = {2012},
month = {oct},
publisher = {IOP Publishing},
volume = {29},
number = {22},
pages = {224005},
author = {Hu, B L and Lin, Shih-Yuin and Louko, Jorma},
title = {Relativistic quantum information in detectors–field interactions},
journal = {Class. Quantum Grav.}
}

@article{Kasprzak2025,
doi = {10.1088/1751-8121/adb66b},
url = {https://doi.org/10.1088/1751-8121/adb66b},
year = {2025},
month = {feb},
volume = {58},
number = {9},
pages = {095301},
author = {Kasprzak, Michael and Tjoa, Erickson},
title = {Transmission of quantum information through quantum fields in curved spacetimes},
journal = {J. Phys. A: Math. Theor.},
}

@article{Sewell1982,
  author = {Geoffrey L. Sewell},
  title = {Quantum fields on manifolds: PCT and gravitationally induced thermal states},
  journal = {Annals of Physics},
  volume = {141},
  number = {2},
  pages = {201--224},
  year = {1982},
  doi = {10.1016/0003-4916(82)90065-1}
}

@article{Wall2012,
  author = {Aron C. Wall},
  title = {Proof of the generalized second law for rapidly changing fields and arbitrary horizon slices},
  journal = {Phys. Rev. D},
  volume = {85},
  number = {6},
  pages = {104049},
  year = {2012},
  doi = {10.1103/PhysRevD.85.104049}
}

@article{Basso2025,
  title={Quantum detailed fluctuation theorem in curved spacetimes: The observer dependent nature of entropy production},
  author={Basso, M. L. W. and Maziero, J. and C\'{e}leri, L. C.},
  journal={Phys. Rev. Lett.},
  volume={134},
  pages={050406},
  year={2025},
  url = {https://doi.org/10.1103/PhysRevLett.134.050406}
}

@article{Araki1976,
  author  = {Araki, Huzihiro},
  title   = {Relative Entropy of States of von Neumann Algebras},
  journal = {Publications of the Research Institute for Mathematical Sciences},
  volume  = {11},
  number  = {3},
  pages   = {809--833},
  year    = {1976},
  doi     = {10.2977/prims/1195191148}
}

@article{Araki1977,
  author  = {Araki, Huzihiro},
  title   = {Relative Entropy for States of von Neumann Algebras {II}},
  journal = {Publications of the Research Institute for Mathematical Sciences},
  volume  = {13},
  number  = {1},
  pages   = {173--192},
  year    = {1977},
  doi     = {10.2977/prims/1195190105}
}

@article{JaksicPillet2014,
  author  = {Jak{\v{s}}i{\'c}, Vojkan and Pillet, Claude-Alain},
  title   = {A Note on the Landauer Principle in Quantum Statistical Mechanics},
  journal = {Journal of Mathematical Physics},
  volume  = {55},
  number  = {7},
  pages   = {075210},
  year    = {2014},
  doi     = {10.1063/1.4884475}
}

@article{Shevchenko2017,
  author  = {Shevchenko, V.},
  title   = {Finite Time Measurements by {Unruh--DeWitt} Detector and
             Landauer's Principle},
  journal = {Annals of Physics},
  volume  = {381},
  pages   = {17--40},
  year    = {2017},
  doi     = {10.1016/j.aop.2017.03.014}
}

\end{document}